\documentclass{article}

\usepackage{amsmath,amssymb,enumerate,bbm,calc,capt-of,ifthen}
\usepackage{epsfig}

\newcounter{cntr}

\newtheorem{theorem}{Theorem}
\newtheorem{definition}[cntr]{Definition}
\newcommand{\dueto}[1]{\textup{\textbf{(#1) }}}
\newtheorem{varremark}[cntr]{Remark}
\newenvironment{remark}{\begin{varremark}\em}{\em\end{varremark}}

\newtheorem{proposition}[cntr]{Proposition}
\newtheorem{lemma}[cntr]{Lemma}

 \newenvironment{proof}{
   \noindent\textbf{Proof.}\ }{\hspace*{\fill}
   \begin{math}\Box\end{math}\medskip}

\newenvironment{proofof}[1]{
  \noindent\textit{Proof of #1.}\ }{\hspace*{\fill}
  \begin{math}\Box\end{math}\medskip}

\newenvironment{proof*}[1]{
  \noindent\textbf{#1\ }}{\hspace*{\fill}
  \begin{math}\Box\end{math}\medskip}

\newcommand{\nin}{\not\in}

\newcommand{\tmfloatcontents}{}
\newlength{\tmfloatwidth}
\newcommand{\tmfloat}[5]{
  \renewcommand{\tmfloatcontents}{#4}
  \setlength{\tmfloatwidth}{\widthof{\tmfloatcontents}+1in}
  \ifthenelse{\equal{#2}{small}}
    {\ifthenelse{\lengthtest{\tmfloatwidth > \linewidth}}
      {\setlength{\tmfloatwidth}{\linewidth}}{}}
    {\setlength{\tmfloatwidth}{\linewidth}}  \begin{minipage}[#1]{\tmfloatwidth}
    \begin{center}
      \tmfloatcontents
      \captionof{#3}{#5}
    \end{center}
  \end{minipage}}
\newcommand{\ETc}{a_0}
\newcommand{\ETv}[1]{a_v(#1)}

\newcommand{\BSLoc}[2]{Y_{#1}(#2)} 
\newcommand{\BSGlob}[3]{\Phi_{#1}(#2,#3)} 
\newcommand{\polepos}[1]{\check{\sigma}_{#1}}

\newcommand{\DLoc}[2]{D_{#1}(#2)}
\newcommand{\DLocCoeff}[2]{D_{#1,#2}}
\newcommand{\DGlob}[3]{\Psi_{#1}(#2,#3)}

\newcommand{\IP}[2]{\langle #1 | #2 \rangle}

\newcommand{\norm}[2]{\left\| #1 \right \|_{#2}}
\newcommand{\abs}[1]{\left| #1 \right|}

\newcommand{\Zak}[0]{\mathcal{Z}}

\newcommand{\rp}[0]{\mathbb{R}^+}

\newcommand{\ninZ}[0]{n \in \mathbb{Z}}

\newcommand{\Mo}[0]{0}
\newcommand{\Nm}[0]{{ }}

\newcommand{\storus}[0]{S^{1}_{\omega}}

\numberwithin{equation}{section}
\numberwithin{cntr}{section}

\begin{document}

\title{Ionization in a 1-Dimensional Dipole Model} 
\author{O. Costin, J.L. Lebowitz and C. Stucchio }


\maketitle

\begin{abstract}
We study the evolution of a one dimensional model atom with $\delta$-function binding potential, subjected to a dipole radiation field $E(t) x$ with $E(t)$ a $2\pi/\omega$-periodic real-valued function. Starting with $\psi(x,t=0)$ an initially localized state and $E(t)$ a trigonometric polynomial, complete ionization occurs; the probability of finding the electron in any fixed region goes to zero.

For $\psi(x,0)$ compactly supported and general periodic fields, we construct a resonance expansion. Each resonance is given explicitly as a Gamow vector, and is $2\pi/\omega$ periodic in time and behaves like the exponentially growing Green's function near $x=\pm \infty$. The remainder is given by an asymptotic power series in $t^{-1/2}$ with coefficients varying with $x$. 
\end{abstract}

\section{Introduction}

The ionization of an atom by an electromagnetic field is one of the central problems of atomic physics. There exists a variety of approximate methods for treating this problem. These include perturbation theory (Fermi's golden rule), numerical integration of the time-dependent Schr\"odinger equation and semi-classical phase space analysis leading to stochastic ionization \cite{cohen:atomPhotonInteractions,chin:multiphoton,graffi:NBodyACStarkExteriorScaling,jensen:stochasticIonizationSurfaces,jensen:semiclassicalStochasticIonization,landaulifshitz}. Rigorous approaches include Floquet theory and complex dilations \cite{graffi:NBodyACStarkExteriorScaling,yajima:ACStarkExteriorScaling,yajima:ACStarkResonances}. Despite this, there are few exact results available for the ionization of a bound particle by a realistic time-periodic electric field of dipole form $\vec{E} ( t ) \cdot \vec{x}$ (an AC-Stark field) for fields of arbitrary strength. The most realistic results we are aware of are based on complex scaling (\cite{graffi:NBodyACStarkExteriorScaling,yajima:ACStarkExteriorScaling,yajima:ACStarkResonances}) and show ionization (for small electric field) of certain bound states of the Coulomb atom as well as defining resonances in some regions of the complex energy plane.

The lack of rigorous results for large electric fields is true not only for  realistic systems with Coulombic binding potential, but even for model systems with short range  binding potentials \cite{cohen:atomPhotonInteractions,chin:multiphoton,geltman:unknown1}. The most idealized version of the latter has an attractive $\delta$-function potential in 1 dimension. The unperturbed Hamiltonian $H_{0}=-\partial_{x}^{2} - 2 \delta(x)$ has a bound state $\phi_{0}(x) = e^{-\abs{x}}$ with energy $-1$, and explicitly known continuum states \cite{costin:cmpionize}. This model has been studied extensively in the literature, but the only rigorous results (known to us) concerning ionization involve short range external forcing potentials rather than dipole interaction; see however \cite{schrader:anastab,MR1618647,MR1447116} for some rigorous bounds on the ionization probability by a dipole potential for finite time pulses. Detailed results for compactly supported forcings were obtained in \cite{costin:jspionize,costin:cmpionize,costin:jpaionize,MR2163573}. In this paper we develop techniques to deal with physically realistic dipole interactions.

We consider the time evolution of a particle in one dimension governed by the Schr\"odinger equation (in appropriate units):
\begin{subequations}
  \label{eq:electricGauge}
  \begin{eqnarray}
    i \partial_t \psi ( x, t ) & = & \left( - \frac{\partial^2}{\partial x^2} - 2 \delta ( x ) \right) \psi ( x, t ) + E ( t ) x \psi ( x, t ) \\
    \psi(x,0) &=& \psi_0(x) \in L^{2}(\mathbb{R})
  \end{eqnarray}
\end{subequations}
Here, $(x,t) \in \mathbb{R} \times \mathbb{R}^+$, $E ( t )$ is real valued, smooth and periodic, $E(t)=E(t+2\pi/\omega)$. We prove that for $E( t )$ a trigonometric polynomial,
\begin{equation}
  E ( t ) = \sum_{n = 1}^N \left( E_n e^{i n \omega t} + \overline{E_n} e^{-i n \omega t} \right),
  \label{eq:defOfE}
\end{equation}
with $N < \infty$ the system always ionizes, i.e. for $\psi ( x, t )$ solving \eqref{eq:electricGauge},
\begin{eqnarray}
  \lim_{t \rightarrow \infty} \int_{-L}^L | \psi ( x, t ) |^2 d x = 0 ,& &\forall L \in \mathbb{R}^+
  \label{eq:introIoniz}
\end{eqnarray}
with the approach to zero at least as fast as $O ( t^{- 1} )$ if $\psi_{0}(x) \in L^{1} \cap L^{2}(\mathbb{R})$. When $E(t)$ is not a trigonometric polynomial (i.e. $N=\infty$ in (\ref{eq:defOfE})), the Floquet Hamiltonian (see below) may have time-dependent bound states and ionization may fail. This is uncommon, but there are examples of time periodic operators where such bound states exist \cite{costin:cmpionize,MR1758988}. 

For general periodic $E(t)$ and for compactly supported initial data, $\psi(x,t)$ can be uniquely decomposed into a sum of (non-$L^{2}$) functions
\begin{equation}
  \psi(x,t) =  \sum_{k=0}^{M-1} \sum_{j=0}^{n_{k}} \alpha_{k,j} t^{j} e^{-i \polepos{k} t} \BSGlob{k,j}{x}{t} + \DGlob{M}{x}{t}
\end{equation}
 (with $1 \leq M < \infty$). The indexing in $k$ is chosen so that $\Im \polepos{k} \leq \Im \polepos{k+1}$. The functions $\BSGlob{k,j}{x}{t}$ are $2\pi / \omega$ periodic in time and exponentially growing in space, behaving near $x=\pm \infty$ like the exponentially growing Green's function of the Floquet Hamiltonian. Note that $\Im \polepos{k} \leq 0$, otherwise the wavefunction would grow exponentially in time. If we define
\begin{equation}
  \label{eq:defgamma}
  \gamma_{k} = - \Im \polepos{k},
\end{equation}
then for small values of $\gamma_{k}$, $2\gamma_{k}$ gives the dominant part of the ionization rate for the $k$-th resonance. The smallest rate, $\gamma_{0}$, gives the overall ionization rate for most experimentally relevant times \cite{cohen:atomPhotonInteractions}. The ionization rate $\gamma_{0}$ will vanish as $E(t) \rightarrow 0$, while $\gamma_{k} \rightarrow \infty$ as $E(t) \rightarrow 0$ for $k \geq 1$. 

The term $\DGlob{M}{x}{t}$ is a remainder after collecting $M$ resonances (provided $\psi(x,t)$ has $M$ resonances), possessing an asymptotic expansion in time with power law terms in $t^{-n/2}$ for $n \geq 1$. $\DGlob{M}{x}{t}$ is computed as the integral of a certain function around a branch cut; the resonant states $\BSGlob{k,j}{x}{t}$ are  the residues of the poles of that same function (this $M$ can not be greater than the number of poles). We note that the polynomially decaying component of the wavefunction, to which we shall refer as the dispersive part, has actually been observed experimentally \cite{rothe:163601}, although under significantly different\footnote{In \cite{rothe:163601}, the authors studied luminescent decay of dissolved organic materials after a pulsed laser excitation.} physical conditions.

Replacing $E(t)$ by $\epsilon E(t)$, $\polepos{0}$ and $\BSGlob{0,0}{x}{t}$ have convergent power series expansions in $\epsilon$ when $\omega^{-1} \nin \mathbb{N}$. When $\epsilon \rightarrow 0$, we have $e^{-i \polepos{0} t} \BSGlob{0,0}{x}{t} \rightarrow e^{i t} e^{-\abs{x}}$, the bound state of $H_0$ and $\DGlob{1}{x}{t}$ goes to the projection of $\psi(x,t)$ on the continuum states of $H_0$. This shows that the first resonance is the analytic continuation in $\epsilon$ of the bound state. This rigorously justifies some standard physics calculations in \cite{gamow:pedestrianintroduction,gamow:first,landaulifshitz} (see also the forthcoming work \cite{strauss:Gamow}, from which we drew inspiration). The Fermi golden rule and multiphoton generalizations can be recovered in our formalism through perturbation theory.
 
All other resonances must come from $\sigma=-i\infty$. That is, as $\epsilon \rightarrow 0$, $\gamma_{k} \rightarrow \infty$, meaning that these states ``ionize instantly'' in the absence of the electric field. We conjecture that such states do not exist for this model. Indeed, in all other cases considered \cite{costin:jspionize,costin:cmpionize}, such states do not exist, but our technique does not rule them out. See Remark \ref{rem:HowToProveOnePole} for more details on this point.

\subsection{Outline of the strategy}

Due to the fact that the binding potential $\delta(x)$ has support $\{0\}$, the behavior of $\psi(0,t)$ and the initial condition completely determine the behavior of the solution. We shall therefore deal mostly with $\psi(0,t)$ which satisfies an autonomous equation, \eqref{eq:floquetDuhamel}. That equation is well posed, and its solution can be extended to $\psi(x,t)$ solving \eqref{eq:electricGauge}. This is sketched in Appendix \ref{sec:wellPosedness}.

Our main tool is the study of the analytic structure of the Zak transform of $\psi(0,t)$ (with $\psi(0,t)=0$ for $t<0$),
\begin{equation}
  \label{eq:zakFirstFound}
  \Zak[\psi(0,\cdot)](\sigma,t) = \sum_{j \in \mathbb{Z}} e^{i \sigma (t+2\pi j/\omega)} \psi(0,t+2\pi j/\omega)
\end{equation}
in the complex $\sigma$ domain. As will be made clear later, see Definition \ref{def:Zak} and Proposition \ref{prop:zakProperties}, it is sufficient to consider the strip $0 \leq \Re \sigma < \omega$, which we shall do henceforth. Unitary evolution of the wavefunction implies that $\Zak[\psi(0,\cdot)](\sigma,t)$ is analytic in $\sigma$ for $\Im \sigma > 0$.

For $E ( t ) = 0$ and $\IP{\psi(x,0)}{e^{-\abs{x}}} \neq 0$, $\Zak[\psi(0,\cdot)](\sigma,t)$  has a pole at $\polepos{k} = -1+\lfloor 1/\omega \rfloor \omega$ corresponding to the eigenvalue $-1$ of $H_{0}$ (with $\lfloor 1/\omega \rfloor$ the largest integer less than $1/\omega$). The residue at the pole is $e^{-i \lfloor 1/\omega \rfloor \omega t}$. If we consider $\sigma$ outside the strip $0 \leq \Re \sigma < \omega$, this pole is repeated at the points $\polepos{k} + m \omega$ (see Definition \ref{def:Zak}, in particular \eqref{eq:zakSigmaQuasiPeriodicity}).

We show that when $E(t) \neq 0$, the poles give rise to the terms $e^{-i \polepos{k} t} \BSGlob{k,j}{x}{t}$, with the residue at the poles corresponding (by a linear transformation) to the Fourier coefficients in time of $\BSGlob{k,j}{x}{t}$. There is also a branch point at $\sigma=0$ which gives rise to the dispersive part of the remainder $\DGlob{M}{x}{t}$.

The proof of complete ionization, \eqref{eq:introIoniz}, involves proving that there do not exist any Floquet bound state (nonzero solution to \eqref{eq:floquet}). This is done by solving the Schr\"odinger equation without the $\delta$-function at zero, and showing that solutions which decay as $x \rightarrow -\infty$ can not be matched continuously at $x=0$ to solutions which decay exponentially as $x \rightarrow +\infty$, implying that $\Im \sigma < 0$.

\subsection{Statement of results}

We consider the Schr\"odinger equation with a time periodic Stark Hamiltonian \eqref{eq:electricGauge} on $\mathbb{R}^{1 + 1}$. $E ( t )$ is continuous, and is given by \eqref{eq:defOfE}. We prove two theorems:

\begin{theorem}
  \label{ionization theorem} {\dueto{Ionization}}Suppose $E(t)$ is a trigonometric polynomial, i.e. $E_n = 0$ for  $n >  N$ (with $N < \infty$). Then for any $\psi_{0}(x) \in L^{2}(\mathbb{R})$ ionization occurs in the sense of (\ref{eq:introIoniz}). If $\psi_{0}(x) \in L^1(\mathbb{R}) \cap L^2(\mathbb{R})$, then the approach to zero is at least as fast as $t^{-1}$.
\end{theorem}

A key tool in proving Theorem \ref{ionization theorem} is the following result on the structure of $\psi(x,t)$. This result holds even if $N=\infty$ in \eqref{eq:defOfE}.

\begin{theorem}
  \label{thm:floquet}
  Suppose $\psi_{0}(x)$ is compactly supported and in $H^{1}$ (finite kinetic   energy). Then, the solution  $\psi(x,t)$ of \eqref{eq:electricGauge}, can be decomposed as:
  \begin{equation}
    \label{eq:psidecomp}
    \psi(x,t) =  \sum_{k=0}^{M} \sum_{j=0}^{n_{k}} \alpha_{k,j} t^{j} e^{-i \polepos{k} t} \BSGlob{k,j}{x}{t} + \DGlob{M}{x}{t}
  \end{equation}
  where $\Im \polepos{k} \leq 0$, $\BSGlob{k,j}{x}{t}$ is $2\pi/\omega$ periodic in time and continuous in $x$. Here $M$ is finite, but the number of resonances may be infinite. The resonant term $\BSGlob{k,j}{x}{t}$ is a Gamow vector and grows like the exponentially large Green's function as $\abs{x} \rightarrow\infty$. $\BSGlob{k,j}{x}{t}$ is continuous in $x$ and $t$, differentiable in $t$ and $x$ except at $x=0$. For $j=0$, $\BSGlob{k,0}{x}{t}$ is an eigenvector of the Floquet Hamiltonian:
  \begin{subequations}
    \label{eq:floquet}
    \begin{equation}
      \left( - i \partial_{t} - \frac{\partial^2}{\partial x^2} - 2 \delta (x) + E ( t ) x \right) \BSGlob{k,0}{x}{t} = \polepos{k} \BSGlob{k,0}{x}{t}
      \label{eq:floquetEqMain}
    \end{equation}
    \begin{equation}
      \BSGlob{k,0}{x}{t} = \BSGlob{k,0}{x}{t+2 \pi / \omega }
      \label{eq:floquetEqTBdry}
    \end{equation}
  \end{subequations}
  
  The remainder $\DGlob{M}{x}{t}$ has the following asymptotic expansion in time:
    \begin{equation}
      \label{eq:dispersivePartBorelSummable}
      \DGlob{M}{x}{t} \sim \sum_{j \in \mathbb{Z}} e^{i j \omega t} 
      \sum_{n = 3}^{\infty} D_{j,n}(x) t^{-n/ 2}
  \end{equation}

  Finally, when $\polepos{k} \in (0,\omega)$ (i.e. $\gamma_{k}=\Im \polepos{k}=0$), then $n_{k}=0$ and $\BSGlob{k,0}{x}{t}$ is an $L^{2}$-eigenvector of the Floquet Hamiltonian and decays with $x$,
  \begin{equation}
    \lim_{x \rightarrow - \infty} \BSGlob{k,0}{x}{t} = \lim_{x \rightarrow \infty} \BSGlob{k,0}{x}{t} = 0 
    \label{eq:floquetEqXBdry} 
  \end{equation}
  In this case, the functions $\BSGlob{k,0}{x}{t}$ and $\DGlob{M}{x}{t}$ are orthogonal.
\end{theorem}

The Gamow vectors described in Theorem \ref{thm:floquet} are time dependent, and can be written explicitly as follows:
\begin{equation}
  \label{eq:explicitFormOfBSGLob}
  \BSGlob{k,j}{x}{t} = \left\{
    \begin{array}{ll}
      \sum_{n} \psi_{n}^{L} e^{i\sqrt{\polepos{k}+n\omega} x} e^{-i n \omega t} e^{-i b(t) x-i a(t)} & x \leq 0\\
      \sum_{n} \psi_{n}^{R} e^{-i\sqrt{\polepos{k}+n\omega} x} e^{-i n \omega t} e^{-i b(t) x-i a(t)} & x \geq 0 \\
    \end{array}
  \right.
\end{equation}
The coefficients $\psi_{n}^{L,R}$ vary with $j$ and $k$. The functions $a(t)$ and $b(t)$ are defined in  \eqref{eq:defOfABC}. It should be noted that if $\Im \polepos{k}=0$, then $\psi_{n}^{R,L} = 0 $ for $n \geq 0$ (otherwise $\BSGlob{k,0}{x}{t}$ would not be in $L^{2}$, and (\ref{eq:floquetEqXBdry}) would be violated).

\begin{remark}
  The PDE \eqref{eq:floquetEqMain} is formally overdetermined if $\Im \polepos{k} = 0$, since it has three boundary conditions (\eqref{eq:floquetEqXBdry} and \eqref{eq:floquetEqTBdry}). This makes nonzero solutions to \eqref{eq:floquet} unlikely, although there may be some special forms of $E ( t )$ for which such a solution can be found. The proof of Theorem  \ref{ionization theorem} is essentially a proof that in the case of $E ( t )$ a trigonometric polynomial, there are no nonzero solutions in this setting. This also implies that $\gamma_{k} > 0$ for all $k$
\end{remark}

\begin{remark} 
  \label{rem:densityArgumentForCompactSupport}
  Although Theorem \ref{thm:floquet} applies only to compactly supported initial conditions in $H^{1}$, if all compactly supported initial conditions ionize, then all initial conditions in $L^{2}(\mathbb{R})$ will ionize. This is a simple application of the following well known result\footnote{A simple proof: for any $\psi$ in $L^{2}(\mathbb{R})$, find a $u$ a distance $\epsilon$ away from $\psi$. Then $\norm{T(t) \psi}{} \leq \norm{T(t)(\psi-u)}{}+\norm{T(t)u}{}$. Note that $\norm{T(t)(u-\psi)}{} \leq \norm{T(t)}{}\epsilon \leq C \epsilon$ (with $C$ the uniform bound on $\norm{T(t)}{}$), and for sufficiently large $t$, $\norm{T(t)u}{} \leq \epsilon$ as well. Thus $\norm{T(t)\psi}{} \leq (C+1)\epsilon$ for sufficiently large $t$.  } to the operator family $T(t) = 1_{[-L,L]}(x) U(t)$ ($U(t)$ is the propagator for \eqref{eq:electricGauge}):

  If $T(t)$ is a uniformly bounded family of bounded operators on $L^{2}(\mathbb{R})$, and if $T(t) u \rightarrow 0$ for $u$ in a dense subset of $L^{2}(\mathbb{R})$, then $T(t) u \rightarrow 0 $ for all $u \in L^{2}(\mathbb{R})$.
\end{remark} 

\begin{remark}
  If $\psi_{0}(x)$ is not compactly supported, then a similar decomposition can be provided, but with extra terms. Essentially, if $\Zak[e^{i \partial_{x}^{2} t} \psi_{0}(x)](\sigma,x,t)$ is a ramified analytic function of $\sigma$, each singularity will lead to a similar singularity in $\Zak[\psi(x,t)](\sigma,x,t)$, leading to other exponential decay terms. Resonances are poles in $\sigma$, the location of which does not depend on the initial condition.
\end{remark}

\begin{remark}
  We believe the dispersive part $\DGlob{M}{x}{t}$ is Borel summable, although this does not follow from our results. To show this, one needs to find exponential bounds on $\Zak[\psi(0,t)](\sigma,t)$ as $\Im \sigma \rightarrow - \infty$, which would also show that there is only one pole, $\polepos{0}$, the analytic continuation of the bound state. 
\end{remark}

\subsection{\label{gauge transforms}Equivalent formulations}

Here we describe some equivalent formulations of \eqref{eq:electricGauge}. This
material is essentially taken from chapter 7 of {\cite{cycon:schrodinger}}. We
will use \eqref{eq:magneticGauge} in the proof of Theorem \ref{ionization
theorem} and \eqref{eq:velocityGauge} in the proof of Theorem \ref{thm:floquet}. We first define some auxiliary functions:
\begin{subequations}
  \label{eq:defOfABC}
  \begin{equation}
    \label{eq:defOfA}
    a ( t ) = \int_0^t b ( s )^2 d s \equiv \ETc t + \ETv{t} 
  \end{equation}
  \begin{equation}
    \label{eq:defOfB}
    b ( t ) = \sum_{n = 1}^{\infty}\left( \frac{E_n}{i n \omega} e^{i n \omega t} + \frac{\bar{E}_n}{- i n \omega} e^{- i n \omega t} \right)
  \end{equation}
  \begin{equation}
    \label{eq:defOfC}
    c ( t ) =  2 \sum_{n = 1}^{\infty} \left( \frac{E_n}{( i n \omega )^2} e^{i n
      \omega t} + \frac{\bar{E}_n}{( - i n \omega )^2} e^{- i n \omega t} \right) \equiv
    \sum_{n = 1}^{\infty}\left( C_n e^{i n \omega t} + \bar{C}_n e^{- i n \omega t} \right)
  \end{equation}
\end{subequations}
where $\ETv{t}$ is $2 \pi / \omega$ periodic and has mean 0, and $\ETc = (\omega/2\pi)\int_0^{2 \pi / \omega} b ( s )^2 d s$. Note that $(1/2)c''(t) = b'(t)=E(t)$. 

Define $\psi_v ( x, t ) \equiv e^{+ i a(t)} e^{+ i b ( t ) (x-c(t))} \psi ( x - c ( t ), t )$; then the following equation for $\psi_v$ is equivalent to  \eqref{eq:electricGauge}:
\begin{equation}
  i \partial_t \psi_v ( x, t ) = \left( - \frac{\partial^2}{\partial x^2} - 2
  \delta ( x - c ( t ) ) \right) \psi_v ( x, t )
  \label{eq:velocityGauge}
\end{equation}
This is the velocity gauge, and the equivalence can be verified by a computation\footnote{Equation \eqref{eq:velocityGauge} differs from what one finds in \cite{cycon:schrodinger}. In \cite{cycon:schrodinger}, the authors take $\tilde{b}(t)=\int_0^t E(s)ds$ and $\tilde{c}(t)=\int_0^t b(t) dt$, which imply that $\tilde{c}(t)=c(t)+c_0+c_v t$. This does not change the essential feature that $(1/2)c''(t) = b'(t)=E(t)$. }. Similarly, there is an equivalent equation in the magnetic gauge. We obtain it by setting $\psi_B ( x, t ) = e^{+i a(t) } e^{+i b(t)x}\psi ( x, t )$: 
\begin{equation}
   i \partial_t \psi_B ( x, t ) = \left( - \frac{\partial^2}{\partial x^2} -
    2 \delta ( x ) + 2i b ( t )\partial_x  \right)\psi_B ( x, t ) \label{eq:magneticGauge}
\end{equation}

\begin{remark}
  Suppose that either $\psi_B(x,t)$ or $\psi_v(x,t)$ are time-periodic solutions of \eqref{eq:magneticGauge} or \eqref{eq:velocityGauge}. Then $\psi(x,t)$ is a time quasi-periodic solution of \eqref{eq:electricGauge}, and $e^{i \ETc t} \psi(x,t)$ is time-periodic.
\end{remark}

Of course, all this follows only after showing that \eqref{eq:electricGauge}, \eqref{eq:velocityGauge} or \eqref{eq:magneticGauge} are well posed. This is discussed in Appendix \ref{sec:wellPosedness}; the basic idea is to solve Duhamel's equation for $\psi(0,t)$, namely \eqref{eq:floquetDuhamel}, and then extend the solution to all $x$.

\subsection{Organization of the paper}

In Section \ref{proof of ionization}, we assume Theorem \ref{thm:floquet} to be true and use it to prove Theorem \ref{ionization theorem}. In Section \ref{floq form} we prove Theorem \ref{thm:floquet}. In Section \ref{conclusion section} we make some concluding remarks, and discuss possible directions of future research. Some technical material is presented in the appendices.

\newcommand{\fbasis}[2]{\varphi_{#1,#2}}
\section{\label{proof of ionization}Ionization}

Based on Theorem \ref{thm:floquet}, we will to show that the Floquet equation \eqref{eq:floquet} in the magnetic gauge has no nonzero solutions with $\Im \sigma=0$ which satisfy \eqref{eq:floquetEqXBdry}. This implies ionization for compactly supported initial data, which by Remark \ref{rem:densityArgumentForCompactSupport} implies ionization for all $\psi_{0}(x) \in L^{2}(\mathbb{R})$.

In Section \ref{ion free}, we solve \eqref{eq:magneticGauge} without a binding potential (the $-2\delta(x)$ term) and characterize the solutions. We then assume a bound state $\BSGlob{k,0}{x}{t}$ exists, expand it in an appropriate basis, and derive necessary conditions on the coefficients to meet the boundary conditions (decay at $x=\pm \infty$ and continuity at $x=0$).

In Section \ref{ion match}, we use the characterization of solutions we constructed in Section \ref{ion free} and show for $E(t)$ a trigonometric polynomial that there are no continuous, nonzero solutions to \eqref{eq:magneticGauge} which vanish at $x=\pm \infty$. The basic technique is to analytically continue, in the $t$ variable, both $\psi_B(0_-,t)$ and $\psi_B(0_+,t)$ (which must coincide) and use the Phragmen-Lindel\"of theorem to show that an associated function must be entire and bounded (and therefore constant). This implies that any localized solution to \eqref{eq:floquet} is zero, and ionization occurs.

\subsection{\label{ion free}Solutions to the free problem}

By Theorem \ref{thm:floquet}, we need to show that \eqref{eq:floquet} has no nontrivial solutions. In the magnetic gauge, this is the same as showing that if $\BSGlob{k,0}{x}{t}$ solves 
\begin{equation}
  \label{eq:floqMag}
  \polepos{k} \BSGlob{k,0}{x}{t} = (-i \partial_t - \partial_x^2 - 2\delta(x)+2ib(t)\partial_x) \BSGlob{k,0}{x}{t}
\end{equation}
with boundary conditions \eqref{eq:floquetEqXBdry} and \eqref{eq:floquetEqTBdry}, then $\BSGlob{k,0}{x}{t} = 0$. 

We begin by solving \eqref{eq:floqMag} without the $\delta$-function binding potential (and letting $\sigma=\polepos{k}$, which causes no confusion in this section),
\begin{equation}
  \sigma \psi ( x, t ) =
  (- i \partial_t - \partial^2_x  + 2 i b ( t )  \partial_x) \psi ( x, t )
  \label{eq:freeMagnetic}
\end{equation}
Taking $\psi(x,t) = e^{\lambda x} \varphi_\lambda (t)$ as an ansatz, we obtain an ODE for $\varphi_\lambda(t)$:
\begin{equation}
  \partial_t \varphi_{\lambda} ( t ) = - i \left( - \sigma  -
  \lambda^2 +2 i \lambda b ( t ) \right) \varphi_{\lambda} ( t )
\end{equation}
This has the following family of solutions (recalling that $c'(t)=2b(t)$):
\begin{eqnarray}
  & \varphi_{\lambda} ( t ) = e^{- i E_{\lambda} t} e^{ \lambda c(t)} &\\
  & E_{\lambda} = - \sigma - \lambda^2 & \nonumber
\end{eqnarray}
To ensure $2 \pi/ \omega$ periodicity in time, we must have $( - \sigma - \lambda^2 ) = m \omega$, $m \in \mathbb{Z}$. This implies that $\lambda = \pm i \sqrt{m \omega + \sigma}$ (with the branch cut of $\sqrt{z}$ taken to be $-i \mathbb{R}^+$). Therefore, \eqref{eq:freeMagnetic} has the family of solutions:
\begin{subequations}
  \begin{eqnarray}
    \label{free solutions}
    \fbasis{m}{\pm}(x,t) & =& e^{\pm \lambda_m x} e^{- i m \omega t}
    e^{\pm \lambda_m c ( t )}\\
    \lambda_m & =&  -i \sqrt{\sigma + m \omega }\label{eq:lambdaDef}
  \end{eqnarray}
\end{subequations}

\subsection{Matching solutions}\label{ion match}

Given the family of solutions to \eqref{eq:freeMagnetic}, we can attempt to solve \eqref{eq:magneticGauge}. Applying Theorem \ref{thm:floquet}, we have three boundary conditions to satisfy:
\begin{subequations}
  \label{eqs:boundaryConditions}
  \begin{equation}
    \BSGlob{k,0}{0}{t}= \BSGlob{k,0}{0_-}{t} = \BSGlob{k,0}{0_+}{t} \label{continuous}
  \end{equation}
  \begin{equation}
    \partial_x \BSGlob{k,0}{0_+}{t} - \partial_x \BSGlob{k,0}{ 0_-}{t } = - 2 \BSGlob{k,0}{0}{t} \label{diff}
  \end{equation}
  \begin{equation}
    \lim_{x \rightarrow \infty }\BSGlob{k,0}{-x}{t} = \lim_{x \rightarrow \infty }\  \BSGlob{k,0}{ + x}{t } = 0 \label{condition at infinity}
  \end{equation}
\end{subequations}

Consider now a solution $\BSGlob{k,0}{x}{t}$. We can expand (formally) $\psi(x,t)$ in terms of the functions $\fbasis{m}{\pm}$ in the regions $x < 0$ and $x > 0$ separately\footnote{The validity of the expansion is proved in Lemma \ref{lem:sidewaysPropagationRieszBasis} in Appendix \ref{sec:proof:lemma:smoothnessOfGreensFunction}.}:

\begin{equation}
  \label{eq:16}
  \BSGlob{k,0}{x}{t} = \left\{ \begin{array}{ll}
    \sum_{m \in \mathbb{Z}} ( \psi^L_{m,+} \fbasis{m}{+} ( x, t ) + \psi^L_{m,-} \fbasis{m}{-} ( x, t ) ), & x \leq 0\\
    \sum_{m \in \mathbb{Z}} ( \psi^R_{m,+} \fbasis{m}{+} (x,t) + \psi^R_{m,-} \fbasis{m}{-} ( x, t ) ),  & x \geq 0
  \end{array} \right.
\end{equation}

For $m \geq \Mo$ (recalling $\polepos{k}\in [0,\omega)$ and examining \eqref{eq:lambdaDef}), the functions $\fbasis{m}{\pm}(x,t)$ are oscillatory in $x$ as $x \rightarrow \pm \infty$. Thus, if the coefficients $\psi^{L,R}_{m,\pm}$ ($m \geq \Mo$) were not zero, then $\BSGlob{k,0}{x}{t}$ would not decay as $x \rightarrow \pm \infty$, violating \eqref{condition at infinity}.

Similarly, we observe that $\fbasis{m}{+}(x,t)$ are exponentially growing when $m < \Mo$ as $x \rightarrow + \infty$, so $\psi^{R}_{m,+}$ must similarly be zero. The same argument applied to the region $x<0$ shows that $\psi^{L}_{m,+}$ must be zero when $m < \Mo$. Therefore after dropping the $\pm$ in the coefficients $\psi^{L,R}_{m,\pm}$, we obtain the result we seek.

Thus, we find that we can actually write $\BSGlob{k,0}{x}{t}$ as:
\begin{equation}
  \label{solution of magnetic problem}
  \BSGlob{k,0}{x}{t}=\left\{ \begin{array}{ll}
      \sum_{m < \Mo} \psi_{m}^L \fbasis{m}{+}(x,t), &  x \leq 0\\
      \sum_{m < \Mo}  \psi_{m}^R \fbasis{m}{-}(x,t), &  x \geq 0
    \end{array}  \right.
\end{equation}
with both sequences $\psi_{m}^{L,R}$ in $l^{2}$. Although this derivation is purely formal, it is proved in Appendix \ref{sec:proof:lemma:smoothnessOfGreensFunction}. It also motivates \eqref{eq:explicitFormOfBSGLob}.

Substituting \eqref{solution of magnetic problem} into the continuity condition \eqref{continuous} yields:
\begin{equation}
  \label{continuity expansion}
  \sum_{m<\Mo} \psi^L_m \Nm e^{-i m \omega t} e^{\lambda_m c(t)} = 
  \sum_{m<\Mo} \psi^R_m \Nm e^{-i m \omega t} e^{-\lambda_m c(t)}
\end{equation}

\begin{proposition}
  \label{rapid decay of coefficients} Suppose $E(t)$ is a trigonometric polynomial with highest mode $N$, that is $E(t)=\sum_{n=1}^N( E_n e^{i n \omega t} + \bar{E}_n e^{-i n \omega t})$. Set $z=e^{-i \omega t}$. Then $\BSGlob{k,0}{0}{t}$ has the decomposition:
  \begin{equation}
    \label{eq:rapidDecayCoefficients}
    \BSGlob{k,0}{0}{t} = f(z) + g(z^{-1})
  \end{equation}
The functions $f(\cdot)$ and $g(\cdot)$ are entire functions of exponential order $2N$, and $g(0)=0$. This shows in particular that $\BSGlob{k,0}{0}{t}$ is continuous. 

The correspondence between $\BSGlob{k,0}{0}{t}$, $f(z)$ and $g(z)$ is as follows. Let $\psi_j$ denote the $j'th$ Fourier coefficient of $\BSGlob{k,0}{0}{t}$, that is $\BSGlob{k,0}{0}{t}=\sum_{j} \psi_j e^{i j \omega t}$. Then letting $f_j$, $g_j$ be the Taylor coefficients of $f(z)$, $g(z)$, we find $f_j=\psi_{-j}$ for $j \geq 0$ and $g_j = \psi_j$ for $j < 0$.
\end{proposition}
The proof of this fact uses results from Section \ref{floq form}, and is deferred to Appendix \ref{sec:proofOfExponentialOrder}. Finally, we state a result we use, proved in most complex analysis textbooks, e.g. \cite{stein:complexanalysis}.
\begin{theorem}
  \label{sec:matching-solutions}
  \dueto{Phragmen-Lindel\"of} Let $f(z)$ be an analytic function of exponential order $2N$, that is $\abs{f(z)} \leq C e^{C' \abs{z}^{2N}}$. Let $S$ be a sector of opening smaller than $\pi/2N$. Then:
  \begin{equation*}
    \sup_{z \in \partial S} \abs{f(z)} \geq \sup_{z \in S} \abs{f(z)}
  \end{equation*}
\end{theorem}

We are now prepared to prove the main result. 

\begin{proofof}{Theorem \ref{ionization theorem}}

  We describe first the case $N=1$ now (i.e. $E(t)=E \cos(\omega t)$; the case of arbitrary $N$ is treated below). The key idea is that we can use \eqref{solution of magnetic problem} to obtain an asymptotic expansion of $\BSGlob{k,0}{0_{+}}{t}$ and $\BSGlob{k,0}{0_{-}}{t}$ in the open right and left half planes in the variable $z=e^{-i\omega t}$ (respectively); to leading order $\BSGlob{k,0}{0_{-}}{t} \sim \psi^{L}_{m} z^{m} e^{ -C \abs{\Re  z}}$ and $\BSGlob{k,0}{0_{+}}{t} \sim \psi^{R}_{m} z^{m} e^{- C \abs{\Re z}}$ (note that $m$ and $C$ may be different). This asymptotic expansion shows that $f(z)$ decays exponentially along any ray $z=r e^{i \phi}$ in the open left or right half planes.

  In fact, the asymptotic expansion allows us to observe that $f(z)$ (the part of $\BSGlob{k,0}{0}{t}$ which is analytic in $z$) must be bounded except possibly on the line $i \mathbb{R}$. Theorem~\ref{sec:matching-solutions} combined with Proposition \ref{rapid decay of coefficients} allow us to conclude that $f(z)$ is bounded on the line $i \mathbb{R}$. This shows $f(z)$ is bounded on $\mathbb{C}$ and hence zero.

  Since $f(z)$ is zero, $\BSGlob{k,0}{0}{t}=g(z) \sim  g_{M }z^{-M}$ for some $M \in \mathbb{N}$ (since $g(z)$ is analytic). But we previously showed also that $\BSGlob{k,0}{0}{t} \sim \psi^{L}_{m} z^{m} e^{ -C \abs{\Re  z}}$. Two asymptotic expansions must agree to leading order; the only way this can happen is if $g(z)=\BSGlob{k,0}{0}{t}=0$.
  
 The main difference between the case $N=1$ (monochromatic field) and $N > 1$ (polychromatic field) is that instead of the exponential asymptotic expansions being valid in the left and right half planes, they are valid in sectors of opening $\pi/N$; to show this we need to apply Theorem \ref{sec:matching-solutions} to the boundaries of these sectors. 

  We now go through the details.
  
  {\it Step 1: Setup}

  Let $\BSGlob{k,0}{x}{t}$ be a solution to \eqref{eq:floquet}. By the hypothesis of Theorem \ref{ionization theorem}, we let $E(t)$ be a nonzero trigonometric polynomial of order $N$. Let $z = e^{-i \omega t}$. Let $\mathfrak{C}(z) = \sum_{j=1}^N \left(\bar{C}_j z^j + C_j z^{-j}\right)$ where the $C_j$ are the coefficients from \eqref{eq:defOfC}. We apply Proposition \ref{rapid decay of coefficients} to $\BSGlob{k,0}{0}{t}$ and  \eqref{continuity expansion} to obtain:
  \begin{multline}
    \label{eq:analyticCont}
    \BSGlob{k,0}{0}{t} = f(z) + g(z^{-1}) \\
    =  \sum_{m < \Mo} \psi^{L}_m  z^{m} e^{+ \lambda_m \mathfrak{C}(z)}
    = \sum_{m < \Mo} \psi^{R}_m  z^{m} e^{- \lambda_m \mathfrak{C}(z)}
  \end{multline}
  The first equality holds by \eqref{eq:rapidDecayCoefficients}, the second by \eqref{solution of magnetic problem} with $x=0$. A priori, equality holds only when $\abs{z}=1$. However, both of the latter two sums are analytic in any neighborhood of the unit circle in which they are uniformly convergent. Thus, $f(z)+g(z^{-1})$ is the analytic continuation of the sum if the sum is convergent in some neighborhood containing part of the unit disk.

  For the rest of this proof, we make the following convention. The functions $\psi^{L,R}(z)$ are defined by
  \begin{subequations}
    \label{eq:psilrDef}
    \begin{equation}
      \psi^L(z) = \sum_{m < \Mo} \psi^{L}_m z^{m} e^{+ \lambda_m \mathfrak{C}(z)}
    \end{equation}
    \begin{equation}
      \psi^{R}(z) = \sum_{m < \Mo} \psi^{R}_m z^{m} e^{- \lambda_m \mathfrak{C}(z)}
    \end{equation}
  \end{subequations}
  for those $z$ for which the sum is convergent. 
  
  {\it Step 2: Convergence of the sum}
  
  We show now that the sum in \eqref{eq:analyticCont} is convergent in a sufficiently large region.
  
  For $\abs{z} \geq 1$ and $\Re \mathfrak{C}(z) > 0$,
  consider the sum $\sum_{m < \Mo} \psi^R_m z^{m} e^{- \lambda_m \mathfrak{C}(z)}$. In this region, since $\Re \mathfrak{C}(z) > 0$, we find that $e^{ \lambda_m \mathfrak{C}(z)} \leq 1$. The coefficients $\psi^{L,R}_m$ are bounded uniformly in $m$ (since they form an $l^2$ sequence). For $\abs{z} > 1$, $z^m$ is geometrically decaying as $m \rightarrow -\infty$. Therefore the series is absolutely convergent when $\abs{z}>1$ and $\Re \mathfrak{C}(z) > 0$.
  
  The same statement holds with $\sum_{m <\Mo} \psi^L_m z^{m} e^{+ \lambda_m \mathfrak{C}(z)}$ in the region where $\Re \mathfrak{C}(z) < 0$.
  
  Let us define the following sets: 
  \begin{equation*}
    S^+ = \textrm{Connected~component~of~} S^{1} \textrm{~in~} \{ z \in \mathbb{C} : |z| \geq 1, \Re \mathfrak{C}(z) > 0 \} 
  \end{equation*}
  \begin{equation*}
    S^{-} = \textrm{Connected~component~of~} S^{1} \textrm{~in~} \{ z \in \mathbb{C} : |z| \geq 1, \Re \mathfrak{C}(z) < 0 \} 
  \end{equation*}
  A plot indicating the structure of these sectors (for a particular choice of $\mathfrak{C}(z)$) is shown in Figure 1 for the case where $N=2$.
  
  By Proposition \ref{rapid decay of coefficients}, we see that $\psi^{R}(z)$ is analytic in $S^+$ and $\psi^{L}(z)$ is analytic in $S^-$, since the sum in \eqref{eq:psilrDef} is convergent there.
  
  We now show that $S^{+}$ and $S^{-}$ must be unbounded since $\mathfrak{C}(z)$ is not constant. First, note that $\mathfrak{C}(z) = \overline{\mathfrak{C}(\bar{z}^{-1})}$. As in the Schwarz reflection principle, define $B=S^{+} \cup (\bar{S}^{+})^{-1}$. Clearly, $\Re \mathfrak{C}(z)=0$ for $z \in \partial_{B}$. If $S^{+}$ is bounded, then $B$ is bounded as well. By the real max modulus principle, $\Re \mathfrak{C}(z)$ must be zero inside $B$, and hence $\Re \mathfrak{C}(z)$ is bounded everywhere, which is impossible.
  
  Finally we show that the regions $S^+$ and $S^-$ ``fill out'' to open sectors as $\abs{z} \rightarrow \infty$. That is to say, if $S$ is some sector in which $\Re z^N > 0$, then for any ray $\{ r e^{i \theta}: r > 1\}$ contained in $S$, there exists $R=R(\theta)$ so that the truncated ray $\{ r e^{i \theta}: r > R(\theta) \} \subset S^+$. 

  Without loss of generality\footnote{
    Suppose $C_N = \rho e^{i \theta}$. Then rather than choosing $z = e^{i \omega t}$, we would substitute $z = e^{i ( \omega t - \theta / N )}$.
  }, let us suppose that $C_N \in \mathbb{R}^+$. For very large $|z|$, we write $\mathfrak{C}(z) = \sum_{j = 1}^N \bar{C}_j z^j + C_j z^{- j} =  \bar{C}_{N} z^N + O ( z^{N - 1} )$. Then setting $z=r e^{i \theta}$, we find that $r^{-N} \mathfrak{C}(r e^{i \theta}) = \bar{C}_N e^{i N \theta} + O(r^{-1})$. Thus, for $r$ sufficiently large and $N \theta \neq (2m+1)\pi/2$, we find that $r^{-N} \mathfrak{C}(r e^{i \theta})$ has either strictly positive real part or strictly negative real part. In particular, if $\abs{N \theta \mp \pi/2} > \epsilon$, then there exists an $R=R(\epsilon,\theta)$ so that $\Re r^{-N} \mathfrak{C}(r e^{i \theta})$ is bounded strictly away from zero.
  
  Motivated by the above, we define the following subsets of $\mathbb{C}$ (with $j=0 \ldots N-1$):
  \begin{subequations}
    \begin{multline}
      A_{j, \epsilon}^+ = \{ r e^{i \theta} : r \geq R(\epsilon,\theta) ,\\
      \theta \in [ - \pi / 2 N + 2 \pi j / N + \epsilon, \pi / 2 N
      + 2 \pi j / N - \epsilon ] \}
    \end{multline}
    \begin{multline}
      A_{j, \epsilon}^- = \{ r e^{i \theta} : r \geq
      R(\epsilon,\theta),\\
      \theta \in [ - \pi / 2 N + 2 \pi ( j + 1 / 2 ) / N +
      \epsilon, \pi / 2 N + 2 \pi ( j + 1 / 2 ) / N - \epsilon ] \}
    \end{multline}
  \end{subequations}
  Clearly, for sufficiently large $R$, $A^+_{j,\epsilon} \setminus B_{R} \subset S^+$ and $A^-_{j,\epsilon} \setminus B_{R} \subset S^-$. Here, $B_{R}$ is the ball of radius $R$ about $z=0$.

  {\it Step 3: Asymptotics of $f(z)$}

  We now show that $f(z) = 0$. We begin by writing $f(z)$ as follows:
 
  \begin{subequations}
    \label{eq:asymptotic f}
    \begin{equation}
      f ( z ) =  \sum_{n=0}^\infty f_n z^n =  - \sum_{n = 1}^{\infty} g_n z^{- n} +
      \sum_{m < \Mo} \psi^R_m \Nm z^m e^{- \lambda_m \mathfrak{C}(z)}, z \in S^+
      \label{rasymptotic f}
    \end{equation}
    \begin{equation}
      f ( z ) = \sum_{n=0}^\infty f_n z^n = - \sum_{n = 1}^{\infty} g_n z^{- n} + 
      \sum_{m < \Mo} \psi^L_m  \Nm z^m e^{+ \lambda_m \mathfrak{C}(z)}, z \in S^-
      \label{lasymptotic f}
    \end{equation}
  \end{subequations}
  
  We let $S_{k}, k = 0, \ldots, 2 N + 1$ be a set of sectors of opening $\pi / ( 2 N + 1 )$ arranged in such a way that the boundaries of $S_k$ avoid the rays $r e^{i \pi ( 2 j + 1 ) / 2 N}$. Therefore, for sufficiently large $|z|$, the boundaries of $S_k$ are contained in either  $A^{+}_{j,\epsilon}$ or $A^{-}_{j,\epsilon}$ except for a compact region. On $\partial S_{k}$, $f(z)$ is decaying as $\abs{z} \rightarrow \infty$, by a simple examination of \eqref{eq:asymptotic f}. Since $f(z)$ is entire (unlike $\psi(z)$), $f(z)$ is also bounded on $\partial S_{k}$ even for small $z$.

  We have shown that $f(z)$ is bounded on $\partial S_k$. Applying the Phragmen-Lindel\"of theorem, $f ( z )$ is therefore bounded on $S_k$. Since $\cup_{k = 0}^{2 n + 1} S_k =\mathbb{C}$, we find $f ( z )$ is constant. Since we know that along any ray contained in  $A_{j, \varepsilon}^{\pm}$, $f ( z )$ is decreasing, we know $f ( z ) = 0$.

  {\it Step 4: Asymptotics of $g(z)$ }

  We now show that $g(z)=0$. We rewrite \eqref{eq:asymptotic f} with $g(z)$ on the left side.
  \begin{subequations}
    \label{eq:3}
    \begin{equation}
      \sum_{n = 1}^{\infty} g_n z^{- n} = \sum_{m < \Mo} \psi^R_m
      \Nm z^m e^{- \lambda_m \mathfrak{C}(z)}, z \in S^+ \label{rasymptotic g}
    \end{equation}
    \begin{equation}
      \sum_{n = 1}^{\infty} g_n z^{- n} = \sum_{m < \Mo} \psi^L_m
      \Nm z^m e^{+ \lambda_m \mathfrak{C}(z)}, z \in S^- \label{lasymptotic g}
    \end{equation}
  \end{subequations}
  
  Since the left sides of \eqref{rasymptotic g} and \eqref{lasymptotic g} are  (convergent) asymptotic power series (for sufficiently large $\abs{z}$), while the right sides of \eqref{rasymptotic g} and \eqref{lasymptotic g} are (convergent) asymptotic series of exponentials, we find that the right side decays much faster than the left side. This is impossible unless both sides are zero.

\end{proofof}

\section{The Floquet Formulation}
\label{floq form}
In this section we prove Theorem \ref{thm:floquet}. To do so we define an auxiliary function $Y(t)= \psi(c(t),t)$ and derive a closed integral equation of Volterra type for it via Duhamel's formula. We then apply the Zak transform in time to the integral equation for $Y(t)$. This yields an integral equation of compact Fredholm type for $\Zak[Y](\sigma,t)$, the Zak transform of $Y(t)$. The integral operator is shown to be analytic in $\sigma$. Applying the analytic Fredholm alternative to this equation, shows that $\Zak[Y](\sigma,t)$ is meromorphic in $\sigma^{1/2}$. The poles corresponds to resonances or bound states, while the branch point corresponds to the dispersive part of the solution.

In Section \ref{sec:timebehaviorAllx} we extend these results from $x=0$ to the entire real line. We show that the wavefunction, considered in the magnetic gauge, can be decomposed in the form \eqref{eq:psidecomp}. 
If $\Im \polepos{k} = 0$, then $\Re \polepos{k} \in (0,\omega)$ and $\BSGlob{k,0}{x}{t}$ corresponds to a Floquet bound state. The remainder $\DGlob{M}{x}{t}$ decays with time, in particular $\norm{\DGlob{M}{x}{t}}{L^\infty} \leq O(t^{-1/2})$ as $t \rightarrow \infty$.

\subsection{\label{sec:setup}Setting up the problem}

Here we work in the velocity gauge and study \eqref{eq:velocityGauge}. Recall that $c(t)$ is $2\pi/\omega$ periodic. We rewrite \eqref{eq:electricGauge} in Duhamel form, using the standard Green's function for the free Schr\"odinger equation, $(4\pi i t)^{-1/2} e^{i x^{2}/4t}$:
\begin{multline}
  \psi_v ( x, t ) = \psi_{v,0} ( x, t ) \\
  +2 i \int_0^t \int_{\mathbb{R}} 
  \exp \left( 
    \frac{i ( x - x' )^2}{4 ( t - t' )} 
  \right) \delta ( x' - c ( t' ) )
  \psi_v ( x', t' ) 
  d x' \frac{d t'}{\sqrt{4 \pi i ( t - t' )}} \label{eq:preFloquetDuhamel}
\end{multline}
where we have defined:
\begin{equation*}
  \psi_{v,0}(x,t)=e^{i \partial_x^2 t} \psi_v (x,0) = \int_{\mathbb{R}} (4\pi i t)^{-1/2} e^{i \abs{x-x'}^{2}/4t} \psi_{v}(x',0) dx'
\end{equation*}
Computing the $x'$ integral explicitly and changing variables to $s = t - t'$, we find:
\begin{multline}
  \label{eq:duhamelGreensFunction}
 \psi_v ( x, t ) = \psi_{v,0}(x,t)\\ + 2 i \int_0^t  \exp \left( \frac{i ( x - c ( t - s ) )^2}{4 s} \right) \psi_v ( c ( t - s ), t - s ) \frac{d s}{\sqrt{4 \pi i s}}
\end{multline}
We now substitute $x = c ( t )$, to obtain a closed equation for $\psi_v ( c ( t ), t )$:
\begin{multline}
  \psi_v ( c ( t ), t ) = \psi_{v, 0} ( c ( t ), t )\\
  + \sqrt{\frac{i}{\pi}}
  \int_0^t \exp \left( \frac{i ( c ( t ) - c ( t - s ))^2}{4 s} \right) \psi_v ( c ( t - s ), t - s ) \frac{d s}{\sqrt{s}} 
\end{multline}

Setting $Y_0(t)=\psi_{v,0}(c(t),t)$ and $Y(t)= \psi(c(t),t)$ we obtain:
\begin{equation}
  Y ( t ) = Y_0 ( t ) + \sqrt{\frac{i}{\pi}} \int_0^t \exp \left( \frac{i ( c( t ) - c ( t - s ) )^2}{4 s} \right) Y ( t - s ) \frac{d s}{\sqrt{s}}
  \label{eq:floquetDuhamel}
\end{equation}

Here the derivation of \eqref{eq:floquetDuhamel} is formal; a sketch of a rigorous derivation can be found in Appendix \ref{sec:wellPosedness}. The basic idea is to solve (\ref{eq:floquetDuhamel}) and then use (\ref{eq:duhamelGreensFunction}) to extend the solution to all $x$. Showing the solution is in $H^{1}$ for each $t$ is accomplished by stationary phase, see Appendix \ref{sec:wellPosedness}.

The main tool of our analysis will be the Zak transform.

\begin{definition}
  \label{def:Zak}
  Let $f(t)=0$ for $t <0$ and $\abs{f(t)} \leq C e^{\alpha t}$ ($\alpha \in \mathbb{R}^{+}$). Then $f(t)$ is said to be Zak transformable. The Zak transform of $f(t)$ is defined (for $\Im \sigma > \alpha$) by:
  \begin{equation}
    \label{eq:defZakTransform}
    \Zak[f](\sigma,t) = \sum_{j \in \mathbb{Z}} e^{i \sigma (t+2\pi j/\omega)} f(t+2\pi j/\omega)
  \end{equation}
  and by the analytic continuation of \eqref{eq:defZakTransform} when $\Im \sigma < \alpha$, provided that the analytic continuation exists (treating $\Zak[f](\sigma,t)$ as a function of $\sigma$ taking values in $L^{2}([0,2\pi/\omega], dt)$).
\end{definition}

\begin{proposition}
  \label{prop:zakProperties}
  $\Zak[f](\sigma,t)$ has the following properties:
  \begin{subequations}
    \begin{equation}
      \label{eq:zakInversion}
      f(t) = \omega^{-1} \int_{i \beta}^{i\beta+\omega} e^{-i \sigma t} \Zak[f](\sigma,t) d\sigma 
    \end{equation}
    If $\Zak[f](\sigma,t)$ is singular for $\Im \sigma = \beta$, this integral is interpreted as the limit of integrals over the contours $[i(\beta+\epsilon), i(\beta+\epsilon)+\omega]$ as $\epsilon \rightarrow 0$ from above.
    \begin{equation}
      \Zak[f](\sigma,t+2\pi/\omega) = \Zak[f](\sigma,t)
    \end{equation}
    \begin{equation}
      \label{eq:zakSigmaQuasiPeriodicity}
      \Zak[f](\sigma+\omega,t) = e^{i \omega t} \Zak[f](\sigma,t)
    \end{equation}
    If $p(t)$ is $2\pi/\omega$-periodic, then:
    \begin{equation}
      \label{eq:CommutesZakPeriodic}
      \Zak[p f](\sigma,t) = p(t) \Zak[f](\sigma,t) 
    \end{equation}
  \end{subequations}
\end{proposition}

With the exception of \eqref{eq:zakInversion}, these results all follow immediately from \eqref{eq:defZakTransform}. See Remark \ref{remark:zakFourierRelation} for an explanation of \eqref{eq:zakInversion}.

\begin{remark}
  Suppose $f(t)$ is Zak transformable, and uniformly bounded in time ($\alpha=0$). Suppose further that the analytic continuation of $\Zak[f](\sigma,t)$ has a singularity (say at $\sigma=0$). Then \eqref{eq:zakSigmaQuasiPeriodicity} still holds, in the sense that for any direction $\theta$, $\Zak[f](\sigma+\omega+0 e^{i \theta},t) = e^{i \omega t} \Zak[f](\sigma+0 e^{i\theta},t)$.
\end{remark}

\begin{remark}
  More information on the Zak transform can be found in, e.g., \cite[p.p. 109-110]{MR1162107}. Our definition differs slightly from that in \cite{MR1162107} by allowing $\sigma$ to take complex values.
\end{remark}

\begin{remark}
  \label{remark:zakFourierRelation}
  One can relate the Zak and Fourier transforms as follows. Let $\hat{f}(k)=\int e^{i k t} f(t) dt$ denote the Fourier transform of $f(t)$. Then: 
  \begin{equation}
    \label{eq:17}
    \Zak[f](\sigma,t) = \frac{\omega}{2\pi} \sum_{n \in \mathbb{Z}} \hat{f}(\sigma+n\omega) e^{-i n \omega t}
  \end{equation}
  The Poisson summation formula, applied to \eqref{eq:defZakTransform}, yields \eqref{eq:17}. Eq. \eqref{eq:zakInversion} follows immediately from \eqref{eq:17}. This relation implies that our approach is equivalent to the Fourier/Laplace transform analysis done in \cite{MR1991533,costin:jspionize,galtbayer:floquetalternative}. The Zak transform is used simply for algebraic convenience.
\end{remark}

We proceed as follows. Applying the Zak transform to \eqref{eq:floquetDuhamel} yields an integral equation of the form
\begin{equation}
  \label{eq:9}
  y(\sigma,t) = y_0(\sigma,t) + K(\sigma) y(\sigma,t)
\end{equation}
with $y(\sigma,t)=\Zak[Y](\sigma,t)$, $y_0(\sigma,t)=\Zak[Y_0](\sigma,t)$  and $K(\sigma)$  the Zak transform of the integral operator in \eqref{eq:floquetDuhamel}. $K(\sigma)$ will be shown to be meromorphic in $\sigma$ as a compact operator family from $L^2(S^1,dt) \rightarrow L^2(S^1,dt)$, except for a branch point at $\sigma=0$.

 We then use the Fredholm alternative theorem to invert $(1-K(\sigma))$. Once this is done, we find:
\begin{equation}
  y(\sigma,t) = (1-K(\sigma))^{-1} y_0(\sigma,t)
\end{equation}
The poles of $(1-K(\sigma))^{-1}$  correspond to resonances, and a branch point at $\sigma=0$ corresponds to the dispersive part of the solution, i.e. the part with polynomial decay in $t^{-1/2}$.

To begin, we determine the analyticity properties of $\Zak[Y_0](\sigma,t)$.

\begin{proposition}
  \label{prop:imSigmaDecay} Suppose $\psi_{0}(x)$ is smooth and compactly supported. Then near $\sigma=0$, $y_0(\sigma,t)$ has the expansion:
  \begin{equation}
    \label{eq:formOfY0}
    \Zak[Y_0](\sigma,t)=y_0(\sigma,t) = \frac{1}{2}\sigma^{-1/2} \int_{\mathbb{R}} \psi_0(x) dx + f(\sigma^{1/2},t)
  \end{equation}
  The function $f(\sigma^{1/2},t)$ is analytic in $\sigma^{1/2}$, and is in $L^2(S^1,dt)$. Also, for some constants $C_{1}$ and $C_{2}$, we have
  \begin{equation*}
    \norm{\Zak[Y_0](\sigma,t)}{L^2(\storus,dt)} \leq C_1 e^{C_2 \abs{\Im \sigma}}
  \end{equation*}
  Here, $\storus$ is the set $[0,2\pi/\omega]$ with periodic boundaries (so that $0=2\pi/\omega$ in $\storus$).

  In fact, the same conclusion follows for $\Zak[\psi(x+c(t),t)](\sigma,t)$.
\end{proposition}

\begin{proof}
  Consider $Y_0(t)=\psi_{v,0}(c(t),t)$ for $t \geq 0$ only (and $Y_{0}(t)=0$ for $t<0$). Then (with slight abuse of notation) we write
  \begin{equation*}
    Y_0(t) = \chi_{\rp}(t) \int_{\mathbb{R}} e^{i k c(t)} e^{i k^2 t}\hat{\psi}_0(k) dk
  \end{equation*}
  Computing the Zak transform yields:
  \begin{multline}
    \label{eq:rndgzc24t}
    \Zak[Y_0](\sigma,t) = \sum_{j \in \mathbb{Z}} e^{i \sigma (t-2\pi j/\omega)} \chi_{\rp}(t-2\pi j/\omega)\int_{\mathbb{R}} e^{i k c(t)} e^{i k^2 (t-2\pi j/\omega)}\hat{\psi}_0(k) dk \\
    = \int_{\mathbb{R}} e^{i k c(t)} \hat{\psi}_0(k)\left[ \sum_{j \in \mathbb{Z}} e^{i \sigma (t-2\pi j/\omega)} e^{i k^2 (t-2\pi j/\omega)} \chi_{\rp}(t-2\pi j/\omega)\right] dk \\
    = \int_{\mathbb{R}} e^{i k c(t)} \hat{\psi}_0(k)\left[ \sum_{n \in \mathbb{Z}} \frac{e^{-i n \omega t}}{i(k^2+\sigma+n\omega)}  \right] dk \\
    = \sigma^{-1/2} (1/2) \int_{\mathbb{R}} \psi_0(y) dy
    + \sigma^{-1/2} (1/2) \int_{\mathbb{R}} (e^{-\sqrt{\sigma}\abs{c(t)-y}}-1) \psi_0(y) dy \\
    + \sum_{n \neq 0} \frac{e^{-i n \omega t}}{2 \sqrt{\sigma+n\omega}} \int_{\mathbb{R}} e^{-\sqrt{\sigma+n\omega}\abs{c(t)-y}} \psi_0(y) dy
  \end{multline}
  The interchange of the sum and integral between lines 1 and 2 is justified (for $\Im \sigma > 0$ and $t$ fixed) since the sum over $j$ is absolutely convergent, as is the integral over $k$. The result is valid for arbitrary $\sigma$ by analytic continuation.

  The change inside the square brackets between lines 2 and 3 comes from the Poisson summation formula in the $t$ variable, and the fact that the Fourier transform of $\chi_{\rp}(t) e^{i(k^2+\sigma)t}$ is $-i (k^2+\sigma+\zeta)^{-1}$(with $\zeta$ dual to $t$).

  The first term on the right side of \eqref{eq:rndgzc24t} agrees with that in \eqref{eq:formOfY0}. Since $(e^{-\sqrt{\sigma}\abs{c(t)-y}}-1)$ is analytic in $\sigma^{1/2}$, the second term is analytic in $\sigma^{1/2}$. The second and third (which is analytic in $\sigma$) terms become $f(\sigma,t)$.

  Since  $\psi_0(x)$ is supported on a compact region, $\abs{c(t)-y}$ is bounded (say by $C_2$) and exponential growth follows.
  
  This result follows for all $x$ rather than simply $0$ simply by translation invariance of $e^{i \partial_{x}^{2} t}$.
\end{proof}

We now determine the Zak transform of the integral operator in \eqref{eq:floquetDuhamel} and compute the resolvent of it.

\subsection{Construction of the resolvent}

\label{sec:constructResolvent}

We now apply the Zak transform to \eqref{eq:floquetDuhamel} to construct an equivalent integral equation.

\begin{proposition}
  \label{prop:PSFLaptFormulaLast}
  Let $f(t)$ be Zak transformable. Consider the integral operator:
  \begin{equation}
    \label{eq:defOfKV}
    K_V f(t) = \sqrt{\frac{i}{\pi}} \int_0^t \exp\left(i \frac{(c(t)-c(t-s))^2}{4s} \right) f(t-s) \frac{ds}{\sqrt{s}}
  \end{equation}
  Then if $\Im \sigma > 0$, we find:
  \begin{multline}
    \Zak[K_V f](\sigma,t) = K(\sigma) f(\sigma,t) \\
    = \sqrt{\frac{i}{\pi}} \int_{0}^{\infty} \exp\left(i \frac{(c(t)-c(t-s))^2}{4s} \right)
    e^{i \sigma s}
    \Zak[f](\sigma,t-s)
    \frac{ds}{\sqrt{s}}
  \end{multline}
\end{proposition}

\begin{proof}
  We rewrite \eqref{eq:defOfKV} as:
  \begin{multline}
    \label{eq:1}
    \sqrt{\frac{i}{\pi}} \int_0^{t} \exp\left(i \frac{(c(t)-c(t-s))^2}{4s} \right) f(t-s) \frac{ds}{\sqrt{s}}\\
    = \sqrt{\frac{i}{\pi}} \int_{\mathbb{R}} \exp\left(i \frac{(c(t)-c(t-s))^2}{4s} \right) f(t-s) \chi_{\rp}(s) \frac{ds}{\sqrt{s}}
  \end{multline}
  Applying $\Zak$ to both sides of \eqref{eq:1} yields
  \begin{multline}
    \Zak[K_V f](\sigma,t) = \sum_{j \in \mathbb{Z}} e^{i \sigma (t+2\pi j/\omega)} [K_V f](t) \\
    = \sqrt{\frac{i}{\pi}} \sum_{j \in \mathbb{Z}} e^{i \sigma (t+2\pi j/\omega)} \int_{\mathbb{R}} \exp\left(i\frac{(c(t)-c(t-s))^2}{4s} \right) f(t+2\pi j/\omega-s) \chi_{\rp}(s)\frac{ds}{\sqrt{s}}\\
    = \sqrt{\frac{i}{\pi}} \int_{\mathbb{R}} \exp\left(i\frac{(c(t)-c(t-s))^2}{4s} \right) e^{i \sigma s} \\
    \times \left[ \sum_{j \in \mathbb{Z}} e^{i \sigma (t-s+2\pi j/\omega)} f(t-s+2\pi j/\omega) \right] \chi_{\rp}(s)\frac{ds}{\sqrt{s}}\\
    = \sqrt{\frac{i}{\pi}} \int_{0}^{\infty} \exp\left(i\frac{(c(t)-c(t-s))^2}{4s} \right) e^{i \sigma s} \Zak[f](\sigma,t-s) \frac{ds}{\sqrt{s}}
  \end{multline}
  This is what we wanted to show.
\end{proof}

We now show that the operator $K(\sigma)$, constructed above, is compact. We decompose $K(\sigma)$ as $K_F(\sigma)+K_L(\sigma)$ (defined shortly), and treat each piece separately. 

\begin{proposition}
  \label{prop:KFanalyticStructure}
  Define $K_F(\sigma):L^2(S^1,dt) \rightarrow L^2(S^1,dt)$ by:
  \begin{equation*}
    K_F(\sigma) f(t) = \sqrt{\frac{i}{\pi}}\int_0^\infty e^{i \sigma s} f(t-s) \frac{ds}{\sqrt{s}}
  \end{equation*}
  Then, $K_F(\sigma)$ is compact and analytic for $\Im \sigma > 0$. It can be analytically continued to $\Im \sigma \leq 0$, $\sigma \neq 0$, and the continuation has a $\sigma^{-1/2}$ branch point at $\sigma=0$.
\end{proposition}
\begin{proof}
  We compute this exactly by expanding $f(t)$ in Fourier series and interchanging the order of summation and integration:
  \begin{equation}
    \label{eq:rndy3egth}
    \sqrt{\frac{i}{\pi}} \sum_{\ninZ} f_n e^{-i n \omega t} \int_0^\infty  e^{i (\sigma+n\omega) s} \frac{ds}{\sqrt{s}} = \sum_{\ninZ} \frac{f_n}{\sqrt{\sigma+n\omega}} e^{-i n \omega t}
  \end{equation}
  This is valid for $\Im \sigma > 0$, as well as $\Im \sigma = 0$ but in this case we must treat the integral as improper.

  Thus, in the basis $e^{-i n \omega t}$, this operator is diagonal
  multiplication by $(\sigma+n\omega)^{-1/2}$. Compactness follows since the
  diagonal elements decay in both directions. Analyticity for $\sigma \neq 0$ follows by inspection of  the right side of \eqref{eq:rndy3egth}, and choosing the branch cut of $\sqrt{\sigma+n\omega}$ to lie on the negative real line.
\end{proof}

\begin{proposition}
  Define $K_L(\sigma):L^2(S^1,dt) \rightarrow L^2(S^1,dt)$ as:
  \begin{multline}
    \label{eq:defOfKUpperHalfPlane}
    K_L(\sigma) f(t) = \sqrt{\frac{i}{\pi}} \int_{0}^{\infty} \left[\exp\left(i \frac{(c(t)-c(t-s))^2}{4s} \right) -1 \right]
    e^{i \sigma s}  f(t-s) \frac{ds}{\sqrt{s}}
  \end{multline}
  Then $K_L(\sigma)$ is compact for $\Im \sigma \geq 0$ and analytic for $\Im
  \sigma > 0$. It has continuous limiting values at $\Im \sigma = 0$. 
\end{proposition}
\begin{proof}
  We rewrite \eqref{eq:defOfKUpperHalfPlane} as:
  \begin{equation}
    \label{eq:ksumform}
    \int_{0}^{2\pi /\omega} \sum_{k=0}^{\infty} \left[\exp\left(i \frac{(c(t)-c(t-s))^2}{4(s+2\pi k/\omega)} \right)-1\right] \frac{e^{i \sigma(s+2\pi k/\omega)}}{\sqrt{s+2\pi k/\omega}} f(t-s) ds
  \end{equation}
  Provided that $\Im \sigma \geq 0$, the sum is decaying at least as fast as $k^{-3/2}$. Each term in the sum is continuous. Thus the sum is absolutely convergent to a smooth function in $t$ and $s$, which is analytic in $\sigma$ (thus the limit is analytic except possibly when $\Im \sigma = 0$). The region of integration is compact, and so is $K_L(\sigma)$. 
\end{proof}

We now analytically continue $K_L(\sigma)$ to the strip $0 < \Re \sigma < \omega$.

\begin{proposition}
  \label{power series trick}
  Let $K'(\sigma)$ be the integral operator defined by:
  \begin{subequations}
    \begin{equation}
      K'(\sigma) f(t) = \int_{0}^{2\pi /\omega}k'_{\sigma}(t,s) f(t-s)ds
    \end{equation}
    \begin{equation}
      \label{eq:kprimedef}
      k'_{\sigma}(t,s) = \frac{\omega}{2\pi i} 
      \int_{\mathcal{C}} \frac{e^{\sigma p}}{1 - e^{\omega p - i \omega s}} \left[ \exp\left( \frac{(c(t)-c(t-s))^2}{4p} \right) -1 \right] \frac{dp}{\sqrt{p}}
    \end{equation}
  \end{subequations}
  where $\mathcal{C}$ is a contour along the real line in the upper half plane which avoids the singularities of the integrand at $p=0$ and $p=i(s+2\pi n/\omega)$ (see the proof for a specific example).

  Then $K'(\sigma)$ is analytic valued operator on $0 < \Re \sigma < \omega$, and vanishes as $\Im \sigma \rightarrow \infty$. For every $\sigma$, $K'(\sigma)$ is compact. Furthermore, $K'(\sigma)$ is the analytic continuation of $K_{L}(\sigma)$.
  
  Finally, for $\sigma=-i \lambda$ (with $\Re \lambda < 0$) or $\sigma=-i \lambda+\omega$, $K(\sigma)$ is analytic in the parameter $\lambda^{1/2}$ or $(\sigma-\omega)^{1/2}$. 
\end{proposition}

\begin{proof}

  {\it Step 1: Analyticity}

  When writing the contour of integration as $\mathbb{R}+0i$, we actually mean the integral along some contour in the upper half plane which avoids the singularities (at $p=0$ and $p=i[s+(2\pi/\omega)n]$) of the integrand but remains close to $\mathbb{R}$. 

  For instance, let $\gamma_{R}(t)=t \omega R/2\pi$ for $t \in \mathbb{R} \setminus [-2\pi/\omega,2\pi/\omega]$, and $\gamma_{R}(t)=Re^{i [\pi-(\omega t + 2\pi)/4]}$ for $t \in [-2\pi/\omega,2\pi/\omega]$. That is, $\gamma_{R}(t)$ travels along the real line, and circles upward around the disk of radius $R$. The integral is then defined as $\lim_{R \rightarrow 0} \int_{\gamma_{R}}\, \cdot \, \, dp$.

  To compute the behavior of the integral, simply take $R=2\pi/\omega$:
  \begin{multline}
    \label{eq:rndy2wrfs}
    \omega^{-1} k'_{\sigma}(t,s) = 
    \int_{\mathbb{R}+0i} \frac{e^{\sigma p}}{1 - e^{\omega p - i \omega s}} \left[\exp\left( \frac{(c(t)-c(t-s))^2}{4p} \right)-1\right] \frac{dp}{\sqrt{p}}\\
    = \int_{\gamma} \frac{e^{\sigma p}}{1 - e^{\omega p - i \omega s}} \left[\exp\left( \frac{(c(t)-c(t-s))^2}{4p} \right)-1\right] \frac{dp}{\sqrt{p}} \\
    + \frac{2\pi i}{\omega} e^{i \sigma s} \left[\exp\left(
      \frac{(c(t)-c(t-s))^2}{4 i s}
    \right)-1\right] \frac{1}{\sqrt{i s}}
  \end{multline}
  The integrand in the first term is analytic since $p$ stays away from $0$ (thus avoiding the essential singularity at $p=0$). It is exponentially decaying both for large positive $p$ (at the rate $e^{(\sigma - \omega)p}$) and for large negative $p$ (at the rate $e^{-\sigma p}$). If $\Re \sigma=0$ or $\Re \sigma =\omega$, the integrand still decays at the rate $p^{-3/2}$, which is integrable.
    
  The last term is singular, but integrable at $s=0$, and analytic elsewhere. Thus, $k'_{\sigma}(t,s)$ has only a singularity of order $s^{-1/2}$, and is analytic elsewhere. This shows that $K'(\sigma)$ is a compact family of operators, analytic on $\sigma$.

  {\it Step 2: Vanishing of the operator as $\Im \sigma \rightarrow +\infty$}

  We examine (\ref{eq:rndy2wrfs}). The first term vanishes as $\Im \sigma \rightarrow  \infty$ by the Riemann-Lebesgue lemma. The second term vanishes since $e^{i \sigma s}$ does. Thus, $k'_{\sigma}(t,s) \rightarrow 0$, and so does $K'(\sigma)$.

  {\it Step 3: Continuation of $K(\sigma)$ }

  To show that $K'(\sigma)=K(\sigma)$ if $\Im \sigma > 0$, we simply move the contour of integration in \eqref{eq:kprimedef} upward and collect residues:
  \begin{multline*}
    \int_{\mathbb{R}+0i} \frac{e^{\sigma p}}{1 - e^{\omega p - i \omega s}} \left[ \exp\left( \frac{(c(t)-c(t-s))^2}{4p} \right)-1\right] \frac{dp}{\sqrt{p}} \\
    = \lim_{N \rightarrow \infty} \Bigg[ 
      \int_{\mathbb{R}+i2\pi N/\omega} \frac{e^{\sigma p}}{1 - e^{\omega p - i \omega s}} \left[\exp\left( \frac{(c(t)-c(t-s))^2}{4p} \right)-1\right] \frac{dp}{\sqrt{p}} \\
      + \sum_{j=0}^N \frac{2\pi i}{\omega} e^{i \sigma (s+2\pi j/\omega)} \left[\exp\left(\frac{(c(t)-c(t-s))^2}{4i (s+2\pi j/\omega)}
    \right)-1\right] \frac{1}{\sqrt{i (s+2\pi j/\omega)}}
    \Bigg] \\
    = \sum_{j=0}^\infty \frac{2\pi i}{\omega} e^{i \sigma (s+2\pi j/\omega)} \left[\exp\left(\frac{(c(t)-c(t-s))^2}{4i (s+2\pi j/\omega)}
    \right)-1\right] \frac{1}{\sqrt{i (s+2\pi j/\omega}}
  \end{multline*}
  
  We then integrate this kernel against an $L^2(S^1, dt)$ function $f(t)$ and obtain:
  \begin{multline*}
    \int_{0}^{2\pi /\omega} \sqrt{\frac{i}{\pi}} \frac{\omega}{2\pi i} \sum_{j=0}^\infty \frac{2\pi i}{\omega} e^{i \sigma (s+2\pi j/\omega)} \\
    \times \left[\exp\left(\frac{(c(t)-c(t-s))^2}{4i (s+2\pi j/\omega)}
    \right)-1\right] 
   \frac{1}{\sqrt{i (s+2\pi j/\omega)}} f(t-s) ds \\
    = \sqrt{\frac{i}{\pi}} \int_0^\infty \left[\exp\left(i \frac{(c(t)-c(t-s))^2}{4s} \right)-1\right]
    e^{i \sigma s}  f(t-s) \frac{ds}{\sqrt{s}}
  \end{multline*}
  This is in agreement with \eqref{eq:defOfKUpperHalfPlane}. Hence, $K'(\sigma)=K_L(\sigma)$ for $\Im \sigma > 0$, $\Re \sigma \in (0,\omega)$ and therefore $K'(\sigma)$ is the analytic continuation of $K_L(\sigma)$.
  
  {\it Step 4: Singularity at $\sigma=0,\omega$}

  We now wish to show that $K'(-i \lambda)$ is analytic in $\sqrt{\lambda}$ for $\sigma=-i \lambda$, and similarly that $K(-i\lambda+\omega)$ is analytic in $\lambda$. To do this, we proceed as in Step 3, but push the contour down instead of up. We rotate the contour $\gamma_{1} \cup \gamma_{2} \cup \gamma_{3}$, with $\gamma_{1} = [-i \infty-R, -R]$, $\gamma_{2}$ which goes around the unit circle of radius $R$ in the upper half plane (as in step 1), and $\gamma_{3}$ which is $[R,R-i \infty]$. This lets us avoid concerning ourselves with the singularities of the integrand; the important behavior is the decay near $p=-i \infty$.

  Note that the integral kernel of $K'(-i \lambda)$ is given by
  \begin{equation*}
    k'_{-i\lambda}(t,s)
    = \int_{ \gamma_{1} \cup \gamma_{2} \cup \gamma_{3} } \frac{e^{-i \lambda p}}{1 - e^{\omega p - i \omega s}} \left[ \exp\left( \frac{(c(t)-c(t-s))^2}{4p} \right)-1\right] \frac{dp}{\sqrt{p}}
  \end{equation*}
  while that of $K'(-i\lambda+\omega)$ is given by:
  \begin{equation*}
    k'_{-i\lambda+\omega}(t,s) = \int_{ \gamma_{1} \cup \gamma_{2} \cup \gamma_{3} } \frac{e^{-i \lambda p} e^{ \omega p}}{1 - e^{\omega p - i \omega s}} \left[ \exp\left( \frac{(c(t)-c(t-s))^2}{4p} \right)-1\right] \frac{dp}{\sqrt{p}}
  \end{equation*}
  First, observe that the integral over $\gamma_{2}$ is analytic in $\lambda$, provided $R \neq \omega s$. Thus, choosing a different $R$ for $\omega s < (3/4)\pi$ and $\omega s > (1/4)\pi$ shows analyticity in $\lambda$.

  We consider the case $\sigma=-i\lambda$, the case $\sigma=-i \lambda+\omega$ being treated similarly. We now observe that, for $\Re p= R$ (the same argument applies to $\Re p = -R$), the integrand (over $\gamma_{3}$ or $\gamma_{1}$) becomes a Laplace transform:
  \begin{multline}
    \label{eq:11}
    \int_{\gamma_{3}}  \frac{e^{-i \lambda p}}{1 - e^{\omega p - i \omega s}} \left[ \exp\left( \frac{(c(t)-c(t-s))^2}{4p} \right)-1\right] \frac{dp}{\sqrt{p}}  \\ 
    = e^{ i \lambda R} \int_{0}^{-i \infty}  \frac{ e^{-i \lambda p}}{1 - e^{\omega (p + R) - i \omega s}} \left[ \exp\left( \frac{(c(t)-c(t-s))^2}{4(p + R)} \right)-1\right] \frac{dp}{\sqrt{p + R}}
  \end{multline}
  
  We then observe that we can rewrite
  \begin{equation*}
    \left[ \exp\left( \frac{(c(t)-c(t-s))^2}{4(p + R)} \right)-1\right] \frac{1}{\sqrt{p + R}} =
    (p+R)^{-3/2} H(c(t),c(t-s),p+R)
  \end{equation*}
  with $H(c(t),c(t-s),p+R)$ analytic in $p$. This follows since $e^{z}-1=O(z)$ near $z=0$. We now substitute this back into \eqref{eq:11} and change variables to $i \lambda p = z$, to obtain:
  \begin{multline}
    \eqref{eq:11} = -i e^{ i \lambda R} \int_{0}^{ \infty} e^{- z} \frac{ 1}{1 - e^{\omega (-iz/\lambda + R) - i \omega s}} \frac{ H(c(t),c(t-s),-iz/\lambda)}{(-i z/\lambda+R)^{3/2}} \frac{dz}{\lambda} \\
    = -i \lambda^{1/2} e^{ i \lambda R} \int_{0}^{ \infty} e^{- z} \frac{ 1 }{1 - e^{\omega (-iz/\lambda + R) - i \omega s}} \frac{ H(c(t),c(t-s),-iz/\lambda)}{(-i z+R \lambda)^{3/2}}   dz
  \end{multline}
  The integrand is analytic in $\lambda$, and absolutely convergent. The power of $\lambda^{1/2}$ makes the net result a ramified analytic function. The same argument can be applied to $\gamma_{1}$, replacing $R$ by $-R$. Thus, we have shown that $k'_{-i\lambda}(t,s)$ is analytic in $\lambda^{1/2}$. This implies that $K(-i \lambda)$ is analytic in $\lambda$. As remarked before, the case $K(-i \lambda+\omega)$ is identical, so the proof is complete.
\end{proof}

Now that it is justified, we can now write $K'=K_L$. In addition, now that $K_L(\sigma)$ and $K_F(\sigma)$ are defined, it is clear that $K_F(\sigma)+K_L(\sigma)=K(\sigma)$. 

Next we show that $K(\sigma)$ grows ad most exponentially as $\Im \sigma \rightarrow \pm \infty$.

\begin{proposition}
  \label{compact vanishing} $K(\sigma)$ vanishes as $\Im \sigma \rightarrow \infty$. 
\end{proposition}
\begin{proof}
  We break $K(\sigma)$ up as $K(\sigma)=K_{F}(\sigma)+K_{L}(\sigma)$. The first term, $K_{F}(\sigma)$ is bounded (away from $\sigma=0$) simply by inspecting \eqref{eq:rndy3egth}. The second vanishes near $\Im \sigma = \infty$ by Proposition \ref{power series trick}.
\end{proof}

We have now shown that $K(\sigma):L^2(\storus,dt) \rightarrow L^2(\storus,dt)$ is an analytic (in $\sigma$) family of compact operators. This allows us to construct the resolvent.

\begin{proposition}
  \label{prop:formOfPoles} The operator $(1-K(\sigma))^{-1}$ is a meromorphic (in $\sigma$) family of bounded operators. This implies that if $(1-K(\sigma))^{-1}$ has a pole of order $n$ at a point $\sigma=\polepos{k}$, we then have the following asymptotic expansion as $\sigma \rightarrow \polepos{k}$:
  \begin{equation}
    \label{eq:meromorphicResolvent}
    (1-K(\sigma))^{-1} = \sum_{j = 0}^{n_{k}} \frac{\BSLoc{k,j}{t} \IP{\BSLoc{k,j}{t}}{\cdot}}{(\sigma-\polepos{k})^{j+1}} + D( \sigma )
  \end{equation}
  where $D(\sigma)$ is analytic near $\polepos{k}$. $\BSLoc{k,j}{t}$ solves $( 1 - K ( \polepos{k} ) )\BSLoc{k,j}{t}= \BSLoc{k,j}{t}$ (with $\BSLoc{k,-1}{t}=0$). The functions $\BSLoc{k,j}{t}$ are all $L^{2}(\storus)$ functions.

  If $\polepos{k}=0$, then the same result holds, except that the poles are in the variable $\sqrt{\sigma}$ instead of $(\sigma-\polepos{k})$.

  An additional result (which we use later) is that $P_{0}y(0,t)= (1/2) \int_{\mathbb{R}} \psi_{0}(x) dx$, where $P_{0}$ is projection onto the zero'th Fourier coefficient.
\end{proposition}

\begin{proof}
  This is merely the analytic Fredholm alternative theorem. There is only one technical point regarding the behavior near $\sigma=0$ due to the fact that $K(\sigma)$ is singular there. 

  This can be remedied as follows. The function $y(\sigma,t)$ satisfies the
  following equation:
   \begin{equation}
     (1-K_F(\sigma)-K_L(\sigma))y(\sigma,t) = y_0(\sigma,t)
   \end{equation}
   We expand $K_F(\sigma) y(\sigma,t)$ as in the proof of Proposition \ref{prop:KFanalyticStructure}. Due to the fact that $K_F(\sigma)$ is singular only in the zero'th Fourier component (see \eqref{eq:rndy3egth}), we find that: 
   \begin{multline*}
     (1-K_F(\sigma)(1-P_0) - K_L(\sigma)) y(\sigma,t) + \sigma^{-1/2} P_0 y(\sigma,t) \\
     = \sigma^{-1/2} (1/2)\int_{\mathbb{R}} \psi_0(x) dx + f(\sigma^{1/2},t)
   \end{multline*}
  
   Here, $P_0$ is the projection onto the zero'th Fourier coefficient of a function. Take as an ansatz that $P_0 y(0,t)=(1/2)\int_{\mathbb{R}} \psi_0(x) dx$. Then, since $K_F(\sigma)(1-P_0) - K_L(\sigma)$ is compact and analytic in $\sigma^{1/2}$, we find that
   \begin{equation*}
     y(\sigma,t) = [1-K_F(\sigma)(1-P_0) - K_L(\sigma)]^{-1}  f(\sigma^{1/2},t)
   \end{equation*}
   is meromorphic in $\sigma^{1/2}$. This implies that our ansatz was
   consistent.
\end{proof}

\begin{proposition}
  \label{prop:yNearZero}
  Define $K_{\epsilon}(\sigma)$ as $K(\sigma)$ with $c(t)$ replaced by $\epsilon c(t)$ (so in particular, $K_{1}(\sigma)=K(\sigma)$). Then the position of the poles of $K_{\epsilon}(\sigma)$ are ramified analytic functions of $\epsilon$, the field strength, except possibly near $\polepos{k}=-i \infty$.  For small $\epsilon$, there is only one pole $\polepos{0}$ near the real axis (corresponding to the dressed bound state), and all other poles are located near $\sigma=-i \infty$.
\end{proposition}
\begin{proof}
  This is basically the analytic implicit function theorem, using the fact that $K_{\epsilon}(\sigma)$ is analytic in $\epsilon$, and $K_{0}(\sigma) = K_{F}(\sigma)$ (c.f. Proposition \ref{prop:KFanalyticStructure}).

  We first show that no poles form spontaneously. Consider a compact set, bounded by the curve $\gamma$. Then define
  \begin{equation*}
    R_{\gamma,\epsilon} = \int_{\gamma} [1-K_{\epsilon}(\sigma)]^{-1} d\sigma
  \end{equation*}
  Provided $[1-K_{\epsilon}(\sigma)]^{-1}$ is analytic on $\gamma$, then $R_{\gamma,\epsilon}$ is analytic in $\epsilon$. For $\epsilon=0$, we find that:
  \begin{equation}
    [1-K_{0}(\sigma)]^{-1}\left(\sum_{n} f_{n} e^{-i n \omega t} \right) = \sum_{n} \left[1-\frac{1}{\sqrt{\sigma+n\omega}} \right]^{-1}f_{n} e^{-i n \omega t}
  \end{equation}
  which has one pole on the real line, and no others. Using the fact that\\ $\norm{K_{\epsilon}(\sigma)-K_{0}(\sigma)}{} \leq C \epsilon$ (with $C$ depending in $\gamma$), we can expand $K_{\epsilon}(\sigma)$ in a Neumann series in $\epsilon$ (provided $\gamma$ excludes the pole of $[1-K(\sigma)]^{-1}$) to show that $[1-K(\sigma)]^{-1}$ is analytic for all small $\epsilon$. Thus $R_{\gamma,\epsilon}=0$ for all small $\gamma$, and is therefore zero when it is analytic. Since $R_{\gamma,\epsilon}$ is analytic until $[1-K_{\epsilon}(\sigma)]^{-1}$ is singular on $\gamma$, we have shown that any poles inside $\gamma$ must have gotten there by crossing $\gamma$. 

  The same argument can be repeated to show that spontaneous poles of higher order do not form, except that we will need to consider $R_{\gamma,\epsilon} = \int_{\gamma} f_{k}(\sigma) [1-K_{\epsilon}(\sigma)]^{-1} d\sigma$ (for $f_{k}(\sigma)$ a function with nonvanishing $k'th$ derivative) instead.

  This further implies that any poles which are not present for $\epsilon=0$ must come from $\sigma=-i\infty$ as $\epsilon$ is ``switched on''. 

  Analyticity of $\polepos{k}$ follows immediately from Theorems 1.7 and 1.8 in \cite[page 368-370]{kato:perturbations} (see also the discussion following Theorem 1.7). These results show that any eigenvalue $\lambda(\epsilon,\sigma)$ of $K_{\epsilon}(\sigma)$ is analytic. Poles occur where $\lambda(\epsilon,\sigma)=1$. By the implicit function theorem, $\polepos{k}=\polepos{k}(\epsilon)$ is ramified analytic.
\end{proof}

\begin{remark}
  \label{rem:HowToProveOnePole}
  If we could compute an upper bound on the norm of $[1-K(\sigma)]^{-1}$ as $\Im \sigma \rightarrow - \infty$, it would be possible to show that $[1-K(\sigma)]^{-1}$ has only one pole: the analytic continuation of the bound state when $E(t)=0$. The upper bound would make it impossible for poles to come from $-i \infty$. In all other cases we have considered \cite{costin:jspionize,costin:cmpionize} it was possible to do this, and we believe our inability to do so here is a technical point rather than anything fundamental.
\end{remark}

\subsection{Time behavior of $\psi(x,t)$}

We have now shown that $K(\sigma)$ is a compact analytic operator. By the Fredholm alternative, $( 1 - K( \sigma ) )^{- 1}$ is a meromorphic operator family. By deforming the contour in \eqref{eq:zakInversion}, we can determine the behavior of $Y ( t )$. Once this is complete, we can calculate $\BSGlob{k,j}{x}{t}$ and $\DGlob{M}{x}{t}$ and finish the proof of Theorem \ref{thm:floquet}.

\begin{proposition}
  \label{poles or decay} The function $Y ( t )$ has the expansion:
  \begin{equation}
    \label{Y asymp} 
    Y(t) = \sum_{k=0}^{M-1} \sum_{j=0}^{n_{k}} \alpha_{k} t^{j} e^{-i \polepos{k} t} \BSLoc{k,j}{t} + \DLoc{M}{t}
  \end{equation}
  with $\BSLoc{k,j}{t}$ the residue at $\polepos{k}$ and $\alpha_{j,k}= (2\pi/\omega)\IP{y_{0}(\polepos{k},t)}{\BSLoc{k,j}{t}} / j!$. $M$ must not be greater than the number of poles of $[1-K(\sigma)]^{-1}$. The function $\DLoc{M}{t}$ has the asymptotic expansion:
  \begin{equation}
    \label{eq:dlocBorelSum}
    \DLoc{M}{t} \sim \sum_{n \in \mathbb{Z}} e^{-i n \omega t} 
    \sum_{j = 3}^{\infty} \DLocCoeff{j}{n} t^{-j/ 2}
  \end{equation}
  The sum over $n$ is convergent in $l^2$, which shows that $\abs{ \DLoc{M}{t} }  = O(t^{-3/2})$.

  Supposing that $\polepos{k} = 0$ for some $k$, $Y(t)= \DLoc{M}{t}$ except that in \eqref{eq:dlocBorelSum} the sum starts at $j=1$ rather than $j=3$.
\end{proposition}

\begin{proof}
  Because $(1-K(\sigma))^{-1}$ is meromorphic in $\sigma$, $y(\sigma,t)$ can be written as
  \begin{equation}
    y(\sigma,t) = (1-K(\sigma))^{-1}y_0(\sigma,t)
    = \sum_{j=0}^{n_{k}} \frac{\BSLoc{k,j}{t} \IP{\BSLoc{k,j}{t}}{y_0(\sigma,t)} }{(\sigma-\sigma_b)^{j}}  + D(\sigma) y_0(\sigma,t)
  \end{equation}
  We compute $Y(t)$ using \eqref{eq:zakInversion}, and shifting the contour:
  \begin{multline}
    \label{eq:rndj234gsfs}
    Y(t) = \omega^{-1} \int_{0_+}^{\omega_-} e^{-i\sigma t} y(\sigma,t) d\sigma\\
    = \omega^{-1}\int_{0_+}^{-i K(M)+0_+} e^{-i\sigma t} y(\sigma,t) d\sigma + \omega^{-1} \int_{-i K(M)}^{-iK(M)+\omega} e^{-i\sigma t} y(\sigma,t) d\sigma\\
    + \omega^{-1} \int_{-i K(M)+\omega_-}^{\omega_-} e^{-i\sigma t} y(\sigma,t) d\sigma + \operatorname{Residues} = \omega^{-1}\int_{-i K(M)}^{-iK(M)+\omega} e^{-i\sigma t} y(\sigma,t) d\sigma \\
    + \omega^{-1}\int_0^{-i K(M)+0} e^{-i\sigma t} y(\sigma+0_+,t)-e^{-(i\sigma+\omega_-)t}y(\sigma+\omega_-,t) d\sigma\\
    + \operatorname{Residues}
  \end{multline}
  If $[1-K(\sigma)]^{-1}$ has more than $M$ poles, then we make $K(M)$ sufficiently large to collect $M$ of them; otherwise, we simply collect all the poles. The residue term is given by:
  \begin{equation*}
    \sum_{k=0}^{M-1} \sum_{j=0}^{n_{k}} \alpha_{k_,j} t^{j} e^{-i \polepos{k} t} \BSLoc{k,j}{t}
  \end{equation*}
  stemming from the $M$ poles with $\Im \polepos{k} > -K(M)$. By \eqref{eq:zakSigmaQuasiPeriodicity}, we can change the integral in the second to last line of \eqref{eq:rndj234gsfs} to:
  \begin{equation}
    \label{eq:rndt2rsgfg}
    \omega^{-1}\int_{0}^{-i K(M)} e^{-i \sigma t}(y(\sigma+0_+,t)-y(\sigma+0_-,t)) d\sigma
  \end{equation}
  Note that $y(\sigma,t)$ is analytic in $\sigma^{1/2}$, and thus $y(\sigma+0_+,t)-y(\sigma+0_-,t)$ can be expanded in a Puiseux series in $\sigma^{1/2}$ (and a Fourier series in $t$). Watson's lemma yields:
  \begin{multline}
    \label{eq:rndzcgwzfdp}
    \eqref{eq:rndt2rsgfg} = \omega^{-1} \int_0^{-i \infty} e^{-i \sigma t} 
    \sum_{n \in \mathbb{Z}} e^{-i n \omega t} \sum_{j=0}^{\infty} D_{j,n} \sigma^{j/2} d\sigma \\
    \sim \omega^{-1} \sum_{n \in \mathbb{Z}} e^{i n \omega t} \sum_{j = 3}^{\infty} D_{j,n}\Gamma(j/2)t^{-j/ 2}
  \end{multline}
  This is what we wanted to show.

  When $\polepos{k}=0$, the result follows simply by noting that the sum over $j$ in \eqref{eq:rndt2rsgfg} starts from $j=-1$ rather than $j=0$, thereby letting the sum on the right of \eqref{eq:rndt2rsgfg} start at $j=1$ instead of $j=3$.
  
  The integral from $-iK(M)$ to $-iK(M)+\omega$ decays at least as fast as $O(e^{-K(M) t})$, and is included in $\DLoc{k}{t}$.
\end{proof}

\label{sec:timebehaviorAllx}

We now reconstruct $\psi(x,t)$ in the velocity gauge. The basic idea is as follows. We know that $\psi_v(c(t),t) = \DLoc{M}{t} + e^{-i \polepos{k} t} \BSLoc{k,j}{t}$. Using the fact that $\delta(x-c(t)) \psi_v(x,t) = \delta(x-c(t)) \psi_v(c(t),t)$, we find that $\psi_v(x,t)$ satisfies the following equation:
\begin{multline*}
  i \partial_t \psi_v(x,t) = -\partial_{x}^{2} \psi_{v}(x,t) - 2\delta(x-c(t))\psi_{v}(x,t) \\
  = -\partial_{x}^{2} \psi_{v}(x,t) -2 \delta(x-c(t)) \psi_{v}(c(t),t)\\
  =  -\partial_{x}^{2} \psi_v(x,t) - 2 \delta(x-c(t))[ \DLoc{M}{t} + e^{-i \polepos{k} t} \BSLoc{k,j}{t}]
\end{multline*}

\begin{proposition}
  \label{prop:borelSummabilityOfPsiOfXAndT}
  The expansion \eqref{eq:psidecomp} holds.
\end{proposition}

We first state a result, proved in Appendix \ref{sec:proof:lemma:smoothnessOfGreensFunction}, which is necessary for the proof.

\begin{proposition}
  \label{prop:greensFunctionConstructed}
  Let $G(\sigma)$ be the Green's function for the equation:
  \begin{equation}
    \label{eq:23}
    (\sigma+i \partial_{t} + \partial_{x}^{2} -  b(t) \partial_{x} ) u(x,t) = -2 \delta(x) f(t)
  \end{equation}
  so that for $\Im \sigma > 0$, $u(x,t)$ decays as $x \rightarrow \pm \infty$. Here, $u(x,t)$ is in $L^{2}(\storus)$ for each $x$, and $f(t) \in L^{2}(\storus)$. Then $G(\sigma)$ can be analytically continued to the region $\Im \sigma \leq 0$. The function $u(x,t)=G(\sigma)[-2 \delta(x) f(t)]$ has the expansion:
  \begin{subequations}
    \begin{equation}
      \label{eq:greensFunctionAppliedToSomething}
      u(x,t) = \left\{
        \begin{array}{ll}
          \sum_{m} u_{m,R} e^{ \lambda_{m,-} x} 2^{-1/2}\lambda_{m,-}^{-1} e^{- i m \omega t} e^{\mp \lambda_{m,-} c ( t )}, & x \geq 0 \\
          \sum_{m} u_{m,R} e^{ \lambda_{m,+} x} 2^{-1/2}\lambda_{m,+}^{-1} e^{- i m \omega t} e^{\mp \lambda_{m,+} c ( t )}, & x \leq 0
        \end{array}
      \right.
    \end{equation}
    \begin{equation}
      \lambda_{m,\pm} = \mp i \sqrt{\sigma+m\omega}
    \end{equation}
  \end{subequations}
  where $f(t) \mapsto \{ u_{m,R}, u_{m,L} \}$ is a mapping from $L^{2}(\storus) \rightarrow l^{2}(\mathbb{Z} \times \{L,R\})$. $G(\sigma)$ is also a continuous map, analytic in $\sigma^{1/2}$ from $L^{2}(\storus) \rightarrow L^{2}(B_{R} \times \storus)$ with $B_{R}=\{ x : \abs{x} < R \}$ for any $R$. Near $\sigma=0$, the we have $G(\sigma) \delta(x) f(t) = \sigma^{-1/2}(1/2) P_{0}f(t) + O(1)$, with $P_{0} f(t)$ the projection onto the zero'th Fourier coefficient of $f(t)$ and the $O(1)$ term being analytic in $\sigma^{1/2}$.
\end{proposition}

\begin{proofof}{Proposition \ref{prop:borelSummabilityOfPsiOfXAndT}}
  We work in the magnetic gauge, to simplify this part of the problem. Note that $\psi_{B}(x,t) = \psi_{v}(x+c(t),t)$, so in particular, $\psi_{B}(0,t)=\psi_{v}(c(t),t)=Y(t)$. Moreover, recall that the Zak transform commutes with periodic operators, such as the coordinate transform $(x,t) \mapsto (x+c(t),t)$.

  Additionally, in what follows, the notation $O(\sigma^{1/2})$ denotes a function analytic in $\sigma^{1/2}$ taking values in $L^{2}(\storus,dt)$.

  By Zak transforming the Schr\"odinger equation in the magnetic gauge, we obtain the following (with $\Psi(\sigma,x,t)=\Zak[\psi](\sigma,x,t)$):
  \begin{multline*}
    (\sigma+i\partial_{t} ) \Psi(\sigma,x,t) - \psi_{0}(x) = \left[-\Delta  + 2 i b(t) \partial_{x} \right] \Psi(\sigma,x,t) -2\delta(x)\Psi(\sigma,0,t) \\
    = \left[-\Delta +2i b(t) \partial_{x} \right]\Psi(\sigma,x,t) -2\delta(x)y(\sigma,t) 
  \end{multline*}
  Bringing all terms besides $-2 \delta(x) y(\sigma,t)$ to the left, the initial condition to the right and inverting the differential operator yields:
  \begin{multline}
    \label{eq:4}
    \Psi(\sigma,x,t) = \left[+\sigma+i\partial_{t} + \Delta  - b(t) \partial_{x} \right]^{-1} \psi_{0}(x)\\
    -  \left[+\sigma+i\partial_{t} + \Delta  - b(t) \partial_{x} \right]^{-1}2\delta(x) y(\sigma,t)
  \end{multline}
  The second term is given by $-G(\sigma)2\delta(x) y(\sigma,t)$. To compute $\psi(x,t)$, we need to invert the Zak transform. Thus:
  \begin{multline}
    \label{eq:5}
    \psi(x,t) = \omega^{-1} \int_{0}^{\omega} \!\!\! e^{-i \sigma t} \Psi(\sigma,x,t) d\sigma \\
    = \int_{0}^{-i K(M)} \!\!\!  \!\!\!\! e^{-i \sigma t} \Psi(\sigma,x,t) d\sigma + \int_{-iM}^{-i K(M)+\omega}  \!\!\!\! \!\!\!\! \!\!\! e^{-i \sigma t} \Psi(\sigma,x,t) d\sigma+\int_{-i K(M)+\omega}^{0}  \!\!\!\! \!\!\!\! e^{-i \sigma t} \Psi(\sigma,x,t) d\sigma \\
    + \textrm{Residues}
  \end{multline}
  
  Note that the first term of (\ref{eq:4}) can be equivalently written as
  \begin{equation*}
    \Zak[(\sigma+i \partial_{t} + \partial_{x}^{2})^{-1}\psi_{0}(x)](x+c(t),t) = \Zak[e^{i \partial_{x}^{2} t} \psi_{0}(x)](\sigma,x+c(t),t)
  \end{equation*}
  By Proposition \ref{prop:imSigmaDecay}, for each $x$, this term takes the form $(1/2)\sigma^{-1/2} \int \psi_{0}(x) dx + f(\sigma^{1/2},t)$ with $f(\sigma^{1/2},t)$ varying with $x$. Thus, the first term is bounded by $C e^{C \abs{\Im \sigma}}$ (see Proposition \ref{prop:imSigmaDecay}). The second is given by $-2G(\sigma) \delta(x) y(\sigma,t)$. The integral along $[-iM, -iM+\omega]$ decays like $O(e^{-M t})$. Thus, (\ref{eq:5}) becomes:
  \begin{multline}
    \label{eq:6}
    \psi(x,t) \\
    = \int_{0}^{-i K(M)} \!\!\!\!\!\! \!\!\!\! e^{-i \sigma t} \Psi(\sigma,x,t) d\sigma - \int_{\omega}^{-i K(M)+\omega}  \!\!\!\!\!\!\!\!\!\!  \!\!\!\! e^{-i \sigma t} \Psi(\sigma,x,t) d\sigma 
    + \textrm{Residues} + O(e^{-M t})
  \end{multline}
  
  We  show that the contour integral in \eqref{eq:6} gives rise to the dispersive part, while residues give rise to the resonance.

  {\bf The Dispersive Part}
  
  To compute the integral term of \eqref{eq:6}, note that we must compute:
  \begin{multline}
    \label{eq:8}
    \DGlob{M}{x}{t} = 
    \int_{0}^{-i K(M)} e^{-i \sigma t} \Psi(\sigma,x,t) d\sigma - \int_{\omega}^{-i K(M)+\omega} e^{-i \sigma t} \Psi(\sigma,x,t) d\sigma =\\
    \int_{0}^{-i K(M)} \!\!\!\!\!\!\!   \!\!\!\! e^{-i \sigma t}  \Zak[e^{i \partial_{x}^{2} t} \psi_{0}(x)](\sigma,x+c(t),t) d\sigma - e^{-i (\sigma+\omega) t} \Zak[e^{i \partial_{x}^{2} t} \psi_{0}(x)](\sigma+\omega,x+c(t),t) d\sigma \\
    - 2 \int_{0}^{-i K(M)}  \!\!\!\! \!\!\!\! e^{-i \sigma t} G(\sigma) \delta(x) y(\sigma,t) - e^{-i (\sigma+\omega)t} G(\sigma+\omega) \delta(x) y(\sigma+\omega,t) d\sigma
  \end{multline}
  Since $\Zak[f](\sigma+\omega,t)=e^{i \omega t} \Zak[f](\sigma,t)$, we find that:
  \begin{multline}
    \label{eq:10}
    e^{-i \sigma t}  \Zak[e^{i \partial_{x}^{2} t} \psi_{0}(x)](\sigma,x+c(t),t) - e^{-i (\sigma+\omega) t} \Zak[e^{i \partial_{x}^{2} t} \psi_{0}(x)](\sigma+\omega,x+c(t),t) \\
    = e^{-i \sigma t}  \Zak[e^{i \partial_{x}^{2} t} \psi_{0}(x)](\sigma,x+c(t),t)
    - e^{-i \sigma t} \Zak[e^{i \partial_{x}^{2} t} \psi_{0}(x)](\sigma-0,x+c(t),t)
  \end{multline}
  Using the fact that $\Zak[e^{i \partial_{x}^{2} t} \psi_{0}(x)](\delta,x+c(t),t)=\sigma^{-1/2} (1/2) \int \psi_{0}(x) dx + O(\sigma^{1/2})$ (by Proposition \ref{prop:imSigmaDecay}) we find that:
  \begin{multline*}
    (\ref{eq:10}) = e^{-i \sigma t } (\sigma+0)^{-1/2} (1/2) \int \psi_{0}(x) dx + O(\sigma^{1/2})\\
    - 
    e^{-i \sigma t } (\sigma-0)^{-1/2} (1/2) \int \psi_{0}(x) dx + O(\sigma^{1/2}) \\
    = e^{-i \sigma t } \sigma^{-1/2} \int \psi_{0}(x) dx + e^{-i \sigma }O(\sigma^{1/2})
  \end{multline*}
  Plugging this into \eqref{eq:8} yields:
  \begin{multline}
    \label{eq:13}
    (\ref{eq:8}) = t^{-1/2} \int \psi_{0}(x) dx + \int_{0}^{-i K(M)} e^{-i \sigma t }O(\sigma^{1/2}) d\sigma \\
    - 2 \int_{0}^{-i K(M)}  \!\!\!\! \!\!\!\! e^{-i \sigma t} G(\sigma) \delta(x) y(\sigma,t) - e^{-i (\sigma+\omega)t} G(\sigma+\omega) \delta(x) y(\sigma+\omega,t) d\sigma
  \end{multline}
  Again using the identity $\Zak[f](\sigma+\omega,t)=e^{i \omega t} \Zak[f](\sigma,t)$, we find:
  \begin{multline}
    \label{eq:14}
    (\ref{eq:13}) = t^{-1/2} \int \psi_{0}(x) dx + \int_{0}^{-i K(M)}  \!\!\!\!  \!\!\!\! e^{-i \sigma t }O(\sigma^{1/2}) d\sigma \\
    - 2 \int_{0}^{-i K(M) }  \!\!\!\! \!\!\!\! e^{-i \sigma t}\left[ G(\sigma+0) \delta(x) y(\sigma+0,t) -  G(\sigma-0) \delta(x) y(\sigma-0,t)\right] d\sigma
  \end{multline}
  Since $G(\sigma) \delta(x)=\sigma^{-1/2} (1/2)P_{0} + O(1)$ near $\sigma=0$, we find the second integral term in (\ref{eq:14}) becomes:
  \begin{multline}
    \label{eq:15}
    \int_{0}^{-i K(M) } e^{-i \sigma t}\Big[ (\sigma+0)^{-1/2}(1/2) P_{0} y(\sigma+0,t) + O(1) y(\sigma+0,t)\\
      -  (\sigma-0)^{-1/2} (1/2)P_{0} y(\sigma,t) + O(1) \delta(x) y(\sigma-0,t) \Big] d\sigma \\
    =  \int_{0}^{-i K(M)} e^{-i \sigma t} \sigma^{-1/2} P_{0} y(0,t) + e^{-i \sigma t}O(\sigma^{1/2}) d\sigma
  \end{multline}
  Using the fact that $P_{0} y(0,t) = (1/2) \int \psi_{0}(x) dx$ (see Proposition \ref{prop:formOfPoles}), and plugging \eqref{eq:15} into (\ref{eq:13}) yields:
  \begin{multline*}
    \DGlob{M}{x}{t} = t^{-1/2} \int \psi_{0}(x) dx + \int_{0}^{-i K(M)} e^{-i \sigma t }O(\sigma^{1/2}) d\sigma \\
    - 2 \int_{0}^{-i K(M)} e^{-i \sigma t}(1/2)\left[\int \psi_{0}(x) dx \right] d\sigma + \int_{0}^{-i K(M)} e^{-i \sigma t } O(\sigma^{1/2}) d\sigma \\
    = \int_{0}^{-i K(M)} e^{-i \sigma t} O(\sigma^{1/2}) d\sigma
  \end{multline*}
  This is the Laplace transform of a function which is analytic in $\sigma^{1/2}$, which by Watson's lemma yields \eqref{eq:dispersivePartBorelSummable}.
 
  {\bf The Residue Term, $\polepos{k} \neq 0$}
  
  By substituting (\ref{eq:meromorphicResolvent}) into \eqref{eq:5}, we find that when $\polepos{k} \neq 0$, the residue term (for each pole) takes the form:
  \begin{equation*}
    -\frac{2}{\omega j!}e^{-i \polepos{k} t} G(\polepos{k}) \delta(x) \BSLoc{k,j}{t} \IP{\BSLoc{k,j}{t}}{y_{0}(\polepos,t)} = \alpha e^{-i \polepos{k} t} \BSGlob{k,j}{x}{t}
  \end{equation*}
  with $\alpha_{k,j}=\IP{\BSLoc{k,j}{t}}{y_{0}(\polepos,t)}$ and $\BSGlob{k,j}{x}{t}=G(\polepos{k}) \delta(x) \BSLoc{k,j}{t}$. Thus, by Proposition \ref{prop:greensFunctionConstructed} (in particular \eqref{eq:greensFunctionAppliedToSomething}), we have proved \eqref{eq:explicitFormOfBSGLob}. 

  This implies that  $\Psi(\sigma,x,t)$ has a pole at $\sigma=\polepos{k}$ with residue $\BSGlob{k,j}{x}{t}$. Since $\Psi(\sigma,0,t)=y(\sigma,t)$, and $y(\sigma,t)$ has a pole at $\sigma=\polepos{k}$ with residue $\BSLoc{k,j}{0}{t}$, we find (equating the poles) that $\BSGlob{k,j}{0}{t}=\BSLoc{k,j}{t}$. Thus, $\delta(x) \BSLoc{k,j}{t} = \delta(x) \BSGlob{k,j}{0}{t}$, which implies that:
  \begin{equation*}
    \BSGlob{k,j}{x}{t} = -2[\polepos{k}+i\partial_{t} + \Delta - b(t) \partial_{x}]^{-1} \delta(x) \BSGlob{kk,j}{0}{t}
  \end{equation*}
  Applying $[\polepos{k}+i\partial_{t} + \Delta - b(t) \partial_{x}]$ to both sides and rearranging yields (\ref{eq:floquet}).

  {\bf The Residue Term, $\polepos=0$}

  Supposing $\polepos{k}=0$ (for some $k$), by Proposition \ref{prop:yNearZero} we find that $y(\sigma,t) = \BSLoc{k,j}{t} \IP{\BSLoc{k,j}{t}}{y_{0}(0,t)} \sigma^{-1/2} + D(\sigma,t) y_{0}(\sigma,t)$. By Proposition \ref{prop:formOfPoles}, we find that $P_{0} y(\sigma,t) = (1/2) \int_{\mathbb{R}} \psi_{0}(x) dx$; thus we find that $P_{0} \BSLoc{k,j}{t} = 0$.

  Since $G(\sigma)=\sigma^{-1/2} P_{0} + O(1)$, and $P_{0} y(0,t)=0$, this means that for small $\sigma$:
  \begin{equation*}
    G(\sigma) \delta(x) \frac{\BSLoc{k,j}{t}}{\sqrt{\sigma}} = \sum_{m \neq 0} Y_{m,L,R}
    \frac{
      e^{\lambda_{m,\mp}x} e^{-i m \omega t} e^{\lambda_{m,\mp} c(t)}
      }{
        \sqrt{\sigma} \sqrt{2} \lambda_{m,\mp}
        }
  \end{equation*}

  The ``residue'' term  therefore becomes:
  \begin{multline}
    - \alpha \frac{2}{\omega} \int_{0}^{-i \infty} e^{-i \sigma t} G(\sigma) \delta(x)\BSLoc{k,j}{t} \frac{d\sigma}{\sqrt{\sigma}} \\
    = -\alpha \frac{2}{\omega} \int_{0}^{-i \infty} e^{-i \sigma t} \sum_{m \neq 0} Y_{m,L,R}
    \frac{
      e^{\lambda_{m,\mp}x} e^{-i m \omega t} e^{\mp \lambda_{m,\mp} c(t)}
    }{
      \sqrt{\sigma} \sqrt{2} \lambda_{m,\mp}
    } d\sigma \\
    = -\sigma \frac{2}{\omega} t^{-1/2} \sum_{m \neq 0} Y_{m,L,R}
    \frac{
      e^{-i \sqrt{m\omega} x} e^{-i m \omega t} e^{\mp i \sqrt{m\omega} c(t)}
    }{
      \mp i\sqrt{2m\omega} 
    } 
    + O(t^{-3/2})
  \end{multline}
  The $O(t^{-3/2})$ term comes from computing a Laplace-like integral of a Puisseux function, and can be incorporated into $\DGlob{M}{x}{t}$. The $O(t^{-1/2})$ term is the zero-energy resonance. This completes the proof.
\end{proofof}

We have thus far proved all of Theorem \ref{thm:floquet} except for (\ref{eq:floquetEqXBdry}).

\begin{proposition}
  \label{prop:zeroCoefficientsOfFloquetBoundState}
  Suppose that $\Im \polepos{k} = 0$. Then (\ref{eq:floquetEqXBdry}) holds and $\psi_{n}^{L,R} = 0$ for all $n < 0$. Furthermore, the pole is of order $1$.
\end{proposition}

\begin{proof}
  It is clear that unitary evolution implies:
  \begin{equation}
    \label{eq:21}
    \frac{\omega}{2\pi}\int_{0}^{2\pi/\omega} \int_{-R}^{R} \abs{ \psi(x,t)}^{2} dx dt \leq 1
  \end{equation}

  If the pole is of order greater than $1$, then:
  \begin{multline*}
    \psi(x,t) = \sum_{j=0}^{n_{k}} t^{j} \alpha_{k,j} e^{-i \polepos{k} t} \BSGlob{k,j}{x}{t}\\
    + \sum_{k' \neq k} t^{j} \alpha_{k',j} e^{-i \polepos{k'} t} \BSGlob{k',j}{x}{t} + \DGlob{M}{x}{t}
  \end{multline*}
  But the second two terms decay, while the first grows with time. This contradicts unitary evolution, unless $n_{k}=0$. Thus the pole must be of first order.

  Now suppose that in the expansion of $\BSGlob{k,0}{x}{t}$, at least one $\psi_{n}^{L,R} \neq 0$ with $n < 0$. Then $\BSGlob{k,0}{x}{t}$ will oscillate with $x$ rather than decay. This implies that:
  \begin{equation*}
    \frac{\omega}{2\pi}\int_{0}^{2\pi/\omega} \int_{-R}^{R} \abs{\BSGlob{k,j}{x}{t}}^{2} dx dt \geq C R
  \end{equation*}
  for sufficiently large $R$ and some $C > 0$. On the other hand, the rest of $\psi(x,t)$ (the dispersive part, and the exponentially decaying poles) which we denote $R(x,t)$ decays with time. This implies that for $t \geq t_{R}$ (with $T_{R}$ chosen large enough so that $\abs{R(x,t)} \leq \epsilon/\sqrt{2R}$ that:
  \begin{multline*}
    \norm{\psi(x,t) }{} = \norm{
      \BSGlob{k,0}{x}{t} + R(x,t)
      }{} 
      \geq 
      \norm{ \BSGlob{k,0}{x}{t} }{} - \norm{ R(x,t) }{}  \geq \sqrt{CR} - \epsilon
  \end{multline*}
  Selecting $R > (2+\epsilon)/C$ causes $\norm{\psi(x,t)}{} \geq 1$, contradicting unitary evolution. 

  Intuitively, what this means is the following. The modes $\psi_{n}^{L,R}$ with $n < 0$ correspond to radiation modes. If such a mode is nonzero, then $\BSGlob{k,0}{x}{t}$ will be emitting ``radiation'' without decaying, which is clearly impossible.
\end{proof}

\section{Concluding Remarks}
\label{conclusion section}

In this paper we studied the interaction of a simple model atom with a dipole radiation field of arbitrary strength. We obtained a resonance expansion, in which resonances can be resolved regardless of their complex quasi-energy. In particular, we obtained a rigorous definition of the ionization rate $\gamma=-2 \Im \polepos{k}$ and Stark-shifted energy, $\Re \polepos{k}$ for the $k$-th resonance. We further showed that complete ionization occurs ($\gamma > 0$) when $E(t)$ is a trigonometric polynomial. 

Some possible future directions of research include:

\subsection{Perturbative and numerical calculations}

The main feature of our method is that it turns a time dependent problem on $\mathbb{R}$ into a compact analytic Fredholm integral equation. This implies that a family of finite dimensional approximations can be used (in the Zak domain) to approximate solutions to the time dependent Schr\"odinger equation. 

We believe that the quasi-energy methodology used here and in related papers \cite{MR2163573,costin:cmpionize,costin:jspionize} can be used for quantitative calculations of realistic physical systems. Perturbative calculations along these lines have recovered Fermi's Golden Rule and the multiphoton effect.

\subsection{Resonance theory}

Significant effort has been devoted to the rigorous definition of resonances and quasimodes. The best results we are aware of are those of \cite{yajima:ACStarkExteriorScaling,yajima:ACStarkResonances}, based on complex scaling, and those based on analytic continuation of the S-matrix, e.g. \cite{MR1191565}. We provide an alternative definition: a quasi-bound state is the coefficient of an exponentially decaying term in the asymptotic expansion for $\psi(x,t)$ near $t=\infty$. We hope to use this definition to provide a more complete picture of the time evolution of $\psi(x,t)$. 

\subsection{Extension to 3 dimensions}
\label{sec:extension3D}
  In the case of $H_0=-\Delta - 2 \delta(\vec{\vec{x}})$ with $\vec{x} \in \mathbb{R}^3$, a similar equation to \eqref{eq:floquetDuhamel} can be derived. Due to the fact that $\delta(\vec{x})$ is not in $H^{-1}(\mathbb{R}^3)$, $\psi(\vec{x},t)$ becomes singular at $t=0^{+}$. This can be remedied by considering weak solutions, and an equation similar in most respects to \eqref{eq:floquetDuhamel} can be derived which governs the evolution \cite{MR1803381}. For this reason, we believe most results can be adapted to this case, as has been done for $H_{0}=-\Delta-2\delta(x)+E(t)\delta(x)$ \cite{costin:cmpionize,MR2163573}.

\appendix

\section{Proof of Proposition \ref{rapid decay of coefficients}}
\label{sec:proofOfExponentialOrder}


We observe that by the results of Section \ref{floq form}, if a bound state exists, then:
  \begin{equation*}
    \psi_B(0,t) = e^{a ( t ) / 4} e^{- i a(t)} \BSLoc{k}{t}
  \end{equation*}
  Setting $z=e^{-i \omega t}$, and $y(z)=\BSLoc{k}{t}$, we wish to show that $y(z)=f(z)+g(z)$ with $f,g$ both entire of exponential order $2n$. This is equivalent to showing that:
  \begin{equation*}
    \abs{\BSLoc{k}{t+i\alpha}} \leq C \exp[C' \exp(\abs{2N \omega \alpha}) ]
  \end{equation*}
  The function $\BSLoc{k}{t}$ satisfies the equation:
  \begin{equation*}
    \BSLoc{k}{t} = \int_{0}^{2\pi /\omega}k'(t,s) \BSLoc{k}{t-s}ds = - \int_{0}^{2\pi /\omega}k'(t,t-s) \BSLoc{k}{s}ds
  \end{equation*}
  with $k'(t,s)$ as defined in \eqref{eq:kprimedef}.
  Thus we obtain the bound:
  \begin{equation}
    \abs{\BSLoc{k}{t+i\alpha}} \leq \int_{0}^{2\pi /\omega}\abs{ k'(t+i\alpha,t+i\alpha-s)} \abs{\BSLoc{k}{s}}ds
  \end{equation}
  and it suffices to bound $\abs{ k'(t+i\alpha,t+i\alpha-s)}$. From the definition of $k'(t,s)$, we find:
  \begin{multline*}
    k'(t+i\alpha,t+i\alpha-s)\\
    = \frac{\omega}{2\pi i} 
    \int_{\mathbb{R}+0i} \frac{e^{\sigma p}}{1 - e^{\omega p+\alpha - i \omega(t-s)}}
    \left[ \exp\left( \frac{
          (c(t+i\alpha)-c(s))^2
        }{4p}
      \right) -1 \right] \frac{dp}{\sqrt{p}}
  \end{multline*}  
  Supposing $\alpha/\omega > 1$ (we are interested in the behavior as $\alpha \rightarrow \infty$), the integrand is analytic for $z=r e^{i \theta}$, $0<r<1$ and $0\leq \theta \leq \pi$. Thus, we can deform the contour from $\mathbb{R}+0i$ to $\gamma=\partial \{z : \Im z < 0 \operatorname{or} \abs{z} < 1 \}$. 

Note that for some constant $C$, $\abs{c(t+i\alpha)} \leq C e^{N\omega \abs{\alpha}}$, since $c(t)$ is a trigonometric polynomial of order $N$. 

We find that there are three regions of integration which contribute to $k'(t+i\alpha,t+i\alpha-s)$. The regions of integration contributing come from the region near $1 - e^{\omega p+\alpha - i \omega(t-s)}=0$ (the pole of the integrand), large $p$ and small $p$.

If the pole is closer to $\mathbb{R}$ than $\pi/\omega$, we deform $\gamma$ up to encircle it at a distance $p i \omega$. Otherwise, we ignore it. Therefore, in any case, for $z \in \gamma$, $1 - e^{\omega p+\alpha - i \omega(t-s)}$ is uniformly bounded away from zero. 

We then split $\gamma=\gamma_{<} \cup \gamma_{>} \cup \gamma_\alpha$ where $\gamma_{<}=\{p \in \gamma: \abs{p} < (Ce^{N\omega \abs{\alpha}} + \norm{c(s)}{L^\infty})^2 \}$ and $\gamma_{>}=\gamma \setminus \gamma_{<}$. We therefore find that:

\begin{multline*}
  \abs{k'(t+i\alpha,t+i\alpha-s)} 
  \leq \abs{\operatorname{residue}} \\
  C \int_{\gamma_{<}} \abs{ \frac{e^{\sigma p}}{1 - e^{\omega p+\alpha - i \omega(t-s)}} 
    \left[ 
      \exp\left( \frac{
          (c(t+i\alpha)-c(s))^2
        }{4p} \right) -1 
    \right] }\frac{dp}{\sqrt{\abs{p}}}\\
  + C \int_{\gamma_{>}} \abs{ \frac{e^{\sigma p}}{1 - e^{\omega p+\alpha - i \omega(t-s)}}
    \left[ \exp\left( \frac{
          (c(t+i\alpha)-c(s))^2
        }{4p} \right) -1 \right]
  } \frac{dp}{\sqrt{\abs{p}}} \\
  \leq C 
\end{multline*}
The residue can be bounded by:
\begin{multline*}
  \abs{\operatorname{residue}} \\ 
  \leq  C \abs{e^{ \sigma (-\alpha+i\omega(t-s))/\omega} \left[ \exp\left( \frac{(c(t+i\alpha)-c(s))^2}{4 (-\alpha+i\omega(t-s))/\omega } \right) -1 \right] \frac{1}{\sqrt{(-\alpha+i\omega(t-s))/\omega}} } \\
  \leq C \exp(C \abs{c(t+i\alpha)}^2) \leq C \exp(C \exp(2N\omega \abs{\alpha}))
\end{multline*}
We bound the integral over the compact region  $\gamma_{<}$ simply by taking absolute values:
\begin{multline*}
  \int_{\gamma_{<}} \abs{ \frac{e^{\sigma p}}{1 - e^{\omega p+\alpha - i \omega(t-s)}} \left[ \exp\left( \frac{(c(t+i\alpha)-c(s))^2}{4p} \right) -1 \right] }\frac{dp}{\sqrt{\abs{p}}} \\
  \leq \abs{\gamma_{<}} C \exp(C \exp(2N \omega \abs{\alpha}))
\end{multline*}
For the integral over $\gamma_{>}$, we use the fact that if $\abs{z} < 1$, $\abs{e^z - 1} \leq e \abs{z}$:
\begin{multline*}
  \int_{\gamma_{>}} \abs{ \frac{e^{\sigma p}}{1 - e^{\omega p+\alpha - i \omega(t-s)}} \left[ \exp\left( \frac{(c(t+i\alpha)-c(s))^2}{4p} \right) -1 \right] } \frac{dp}{\sqrt{\abs{p}}} \\
  \int_{\gamma_{>}} \abs{ \frac{e^{\sigma p}}{1 - e^{\omega p+\alpha - i \omega(t-s)}} \frac{ (Ce^{N\omega \abs{\alpha}} + \norm{c(s)}{L^\infty})^2  } {\abs{p}} } \frac{dp}{\sqrt{\abs{p}}} \\
  \leq C e^{2N \omega \abs{\alpha}} \int_{\gamma_{>}} \abs{ \frac{e^{\sigma p}}{1 - e^{\omega p+\alpha - i \omega(t-s)}} p^{-3/2}  } dp \leq C \exp(C \exp(2N \omega \abs{\alpha}))
\end{multline*}
Combining these estimates, we find that $k'(t+i\alpha,t+i\alpha-s)$ has the required growth as $\alpha \rightarrow \infty$, hence $\BSLoc{k}{t}$ does. The same argument applies as $\alpha \rightarrow -\infty$.

\section{Proof of Proposition \ref{prop:greensFunctionConstructed}}
\label{sec:proof:lemma:smoothnessOfGreensFunction}

\newcommand{\hil}{\mathcal{H}}
\newcommand{\psicoeff}[2]{\psi_{#1,#2}}
\newcommand{\hbasis}[2]{\phi_{#1,#2}}
\newcommand{\hbasisz}[3]{\phi_{#1,#2,#3}}

We state a few results we need.

\begin{theorem}
  \label{thm:KatoEigenvector}
  \dueto{T. Kato, \cite[page 368]{kato:perturbations}}
  If a family $T(\sigma)$ of closed operators on $X$ depending on $\sigma$ holomorphically has a spectrum consisting of two separated parts, the subspaces of $X$ corresponding to the separated parts also depend on $\sigma$ holomorphically.
\end{theorem}

\begin{remark}
  \label{remark:katoEigenvector}
  A few words of explanation are in order. In \cite{kato:perturbations}, they are given in the commentary following the theorem.

  The analytic dependence of the separated parts of the spectrum means the following. Let $M_\sigma$, $M'_\sigma$ be the spectral subspaces of $T(\sigma)$, related to the two separated parts. Then there exists an analytic function $U(\sigma)$ (called the transformation function), with analytic inverse, so that $M_\sigma=U(\sigma) M_0$ and $M'_\sigma = U(\sigma) M'_0$. For fixed $\sigma$, both $U(\sigma)$ and $U^{-1}(\sigma)$ are bounded operators on the Hilbert space.

  In addition, the spectral projections $P_M(\sigma)$ and $P_{M'}(\sigma)$ can be written as:
  \begin{subequations}
    \label{eq:spectralProjectionsOfSeparatedParts}
    \begin{eqnarray}
      P_M(\sigma)&=&U(\sigma) P_M(0) U^{-1}(\sigma)\\
      P_M(\sigma)&=&U(\sigma) P_{M'}(0) U^{-1}(\sigma)
    \end{eqnarray}
  \end{subequations}
\end{remark}


We now prove a Lemma which allows us to reconstruct $\Psi(\sigma,x,t)$ given solely information about $\Psi(\sigma,0,t)$. The basic idea is to treat the Schr\"odinger equation as an evolution equation in $x$, with a ``Hamiltonian'' that is periodic in $t$.

\begin{lemma}
  \label{lem:sidewaysPropagationRieszBasis}
  Define the Hilbert space $\hil = A(\storus,d t) \oplus L^2(\storus,dt)$, with $A(\storus, dt)$ defined by the norm:
  \begin{equation}
    \norm{f(t)}{A}^{2}=\IP{f(t)}{\abs{\sigma+i \partial_{t}} f(t)}_{L^{2}(\storus,dt)}
  \end{equation}
  Note that the norm on $A(\storus,dt)$ is equivalent to the norm on $H^{1/2}(\storus,dt)$ except at $\sigma=0$. Then there exists a sequence $N_{m}$ with $0 < \inf_{m} \abs{N_{m}} \leq \sup_{m} \abs{N_{m}}$ so that if we define $\phi_{m,\pm}$ and $\lambda_{m,\pm}$ as follows,
  \begin{subequations}
    \label{eq:defOfphimplusminusAndLambdaInAppendix}
    \begin{equation}
      \label{eq:defOfphimplusminusInAppendix}
      \phi_{m,\pm} = N_m
      \left(
        \begin{array}{c}
          2^{-1/2}\lambda_{m,\pm}^{-1} e^{- i m \omega t} e^{\mp \lambda_{m,\pm} c ( t )}\\
          2^{-1/2} e^{- i m \omega t} e^{\mp \lambda_{m,\pm} c ( t )}
        \end{array}
      \right) 
    \end{equation}
    \begin{equation}
      \label{eq:defOfLambdaInAppendix}
      \lambda_{m,\pm} = \mp i \sqrt{\sigma + m \omega}
    \end{equation}
  \end{subequations}
  then $\phi_{m,\pm}$ is a Riesz basis for $\hil$. Furthermore:
  \begin{equation*}
    H = \left[
      \left(\begin{array}{cc}
          0 & 1\\
          \sigma + i \partial_t & 0
        \end{array}\right)
      +
      \left(\begin{array}{cc}
          0 & 0\\
          0 & b ( t )
        \end{array}\right)
    \right]
  \end{equation*}
  is diagonal in this basis, with $H \phi_{m,\pm} = \lambda_{m,\pm} \phi_{m,\pm}$. 

  Moreover, if we define $\hil^{+}$ as the span of $\{ \phi_{m,+} \}_{m \in \mathbb{Z}}$ and $\hil^{-}$ as the span of $\{ \phi_{m,-} \}_{m \in \mathbb{Z}}$ then $e^{x H}$ is defined, bounded and analytic in $\sigma$ on $\hil^{+}$ for $x \leq 0$, and on $\hil^{-}$ for $x \geq 0$.
\end{lemma}

\begin{remark}
  Note that the norm on $A(\storus,dt)$ controls the $\lambda^{-1}_{0,\pm}$ term in the first component of $\phi_{0,\pm}$, which would otherwise blow up as $\sigma \rightarrow 0$. 
\end{remark}

We are nearly ready to prove Lemma \ref{lem:decayOfSqrtSigmaPlusNOmega}. First a minor technical point.

\begin{remark}
  \label{lem:decayOfSqrtSigmaPlusNOmega}
  Consider the sequence $\sqrt{\sigma+n\omega}$, with $\Re \sigma \in (0,\omega)$. For $n$ negative, $\Im \sqrt{\sigma+n\omega} $ grows like $\sqrt{\abs{n}}$. For $n$ positive, $\Im \sqrt{\sigma+n\omega} = O(n^{-1/2})$, and is uniformly bounded below.
\end{remark}

\begin{proofof}{Lemma \ref{lem:sidewaysPropagationRieszBasis}}
  It is a simple calculation to show $\phi_{m,\pm}$ are eigenvectors of $H$ with eigenvalues $\lambda_{m,\pm}$. To show that $\{ \phi_m^{\pm} \}$ is a Riesz basis for $\hil$, we  show that $H$ is a bounded perturbation of a normal operator. Consider the family of operators (analytic in $\zeta$) on $\hil$:
  \[ H_{\zeta} \left(\begin{array}{c}
      u\\
      u_x
    \end{array}\right) = \left(\begin{array}{cc}
      0 & 1\\
      \sigma + i \partial_t & 0
    \end{array}\right) \left(\begin{array}{c}
      u\\
      u_x
    \end{array}\right) + \left(\begin{array}{cc}
      0 & 0\\
      0 & \zeta b ( t )
    \end{array}\right) \left(\begin{array}{c}
      u\\
      u_x
    \end{array}\right) \]
  Consider also the family of vectors (parameterized by $\zeta$):
  \begin{eqnarray*}
    \{ \phi_{m, \zeta}^{\pm} \} &=& \left\{ N_{m,\zeta} \left(\begin{array}{c}
          2^{-1/2}\lambda_{m,\pm}^{-1} e^{- i m \omega t} e^{ \zeta \lambda_{m, \pm} c ( t )}\\
          2^{-1/2} e^{- i m \omega t} e^{\mp \zeta \lambda_{m,\pm} c ( t )}
        \end{array}\right) \right\}_{m \in \mathbb{Z}}
  \end{eqnarray*}

  $N_{m,\zeta}$ is a normalizing constant which is defined implicitly; we discuss it below. For $\zeta=0$, $N_{m,\zeta}=1$.

  A simple calculation shows that $( \phi_{m, \zeta}^{\pm}, \lambda_{m,\pm} )$ are eigenvector/eigenvalue pairs of $H_{\zeta}$. In particular, each $\lambda_{m,\pm}$ is separate from all the others. For $\zeta=0$, they are also orthonormal in $\hil$. Let $P^\pm_m(\zeta)$ be the associated spectral projection operators, given by
  \begin{equation*}
    P^\pm_m(\zeta) = \int_{\gamma_m} (H_\zeta - z)^{-1} d z
  \end{equation*}
  where $\gamma_m$ is a closed curve containing only $\lambda^\pm_{m,\zeta}$, and no other eigenvalue of $H_\zeta$.
  
  Let $U(\zeta)$ be the transformation function of Theorem \ref{thm:KatoEigenvector} (on page \pageref{thm:KatoEigenvector}, see also Remark \ref{remark:katoEigenvector} and \eqref{eq:spectralProjectionsOfSeparatedParts}). Since each eigenvalue $\lambda_{m,\pm}$ is separated from all the others and varies analytically (except near $\sigma=0$), Theorem \ref{thm:KatoEigenvector} implies that:
  \begin{equation*}
    P^\pm_m(\zeta) = U(\zeta) P^\pm_m(0) U^{-1}(\zeta)
  \end{equation*}
  Note now that $P^\pm_m(0)=\IP{{ }\cdot { }}{\phi^\pm_{m,0}} \phi^\pm_{m,0}$. Therefore, we can write:
  \begin{equation*}
    P^\pm_m(\zeta) = \IP{U(\zeta)^{-1} { } \cdot { }}{\phi^\pm_{m,0}} U(\zeta) \phi^\pm_{m,0} = \IP{ { } \cdot { }}{[U(\zeta)^{-1}]^\ast \phi^\pm_{m,0}} U(\zeta) \phi^\pm_{m,0}
  \end{equation*}
  We know that $U(\zeta) \phi^\pm_{m,0}$ is a vector in the direction
  \begin{equation*}
    \left(\begin{array}{c}
        2^{-1/2}\lambda_{m,\pm}^{-1} e^{- i m \omega t} e^{\mp \zeta \lambda_{m, \pm} c ( t )}\\
        2^{-1/2} e^{- i m \omega t} e^{\mp \zeta \lambda_{m,\pm} c ( t )}
      \end{array}\right) 
  \end{equation*}
  but this determines $U(\zeta) \phi^\pm_{m,0}$ only up to a constant. $N_{m,\zeta}$ denotes this constant (see \eqref{eq:spectralProjectionsOfSeparatedParts} in Remark \ref{remark:katoEigenvector}). Because $U(\zeta)$ is bounded above and below, we find that that $0 < \norm{U(\zeta)^{-1}}{}^{-1} \leq \abs{N_{m,\zeta}} \leq \norm{U(\zeta) }{}$. Note also that $N_{m,\zeta}$ is not necessarily real.

  To compute the expansion of a function $\psi(t)$ in this basis, we use the formula $\psi^\pm_m = \IP{U(1)^{-1} \psi}{\phi^\pm_{m,0}}$. Since $\phi^\pm_{m,0}$ is an orthonormal basis, this set of coefficients is clearly in $l^2$, with $l^2$ norm bounded below by $\norm{U(1)}{}^{-1} \norm{\psi(0,t)}{\hil}$ and above by $\norm{U(1)^{-1}}{} \norm{\psi(0,t)}{\hil}$. 
  
  Finally, we need to show that $\sum_{m} P^+_m(\zeta) + P^-_m(\zeta) = 1$, where the sum is interpreted to converge in the strong topology. The sum is strongly convergent when $\zeta=0$, since the $P^\pm_m(0)$ are orthogonal projections. Now multiply  on the left and right by $U(\zeta)$ and $U(\zeta)^{-1}$ (recalling Remark \ref{remark:katoEigenvector} and \eqref{eq:spectralProjectionsOfSeparatedParts}), which are continuous operators (in norm and therefore in the strong topology):
\begin{multline*}
  1 = U(\zeta) U(\zeta)^{-1} = U(\zeta) \left(\sum_{m} P^+_m(0) + P^-_m(0) \right) U(\zeta)^{-1}\\
  = \sum_{m}U(\zeta)[P^+_m(0) + P^-_m(0)]U(\zeta)^{-1} = \sum_{m} P^+_m(\zeta) + P^-_m(\zeta)
\end{multline*}
This completes the proof of the Riesz basis property. To show boundedness of $e^{x H}$, simply note that the real part of the eigenvalues of $H$ is bounded above on $\hil^{+}$ and bounded below on $\hil^{-}$ (though not uniformly in $\sigma$). Thus, $e^{x H}$ is bounded on $\hil^{+}$. Analyticity follows simply by observing that the eigenvalues and eigenfunctions are analytic in $\sigma^{1/2}$.
\end{proofof}

We are now prepared to prove Proposition \ref{prop:greensFunctionConstructed}.

\begin{proofof}{Proposition \ref{prop:greensFunctionConstructed}}
  Note that (\ref{eq:23}) can be rewritten as:
  \begin{equation*}
    \partial_{x} 
    \left(
      \begin{array}{c}
        u\\
        u_x
      \end{array}
    \right)
    = H 
    \left(
      \begin{array}{c}
        u\\
        u_x
      \end{array}
    \right)
  \end{equation*}
  Away from $x=0$, the solution $u(x,t)$ can be written (formally) as:
  \begin{equation}
    \left(
      \begin{array}{r}
        u(x,t)\\
        \partial_{x} u(x,t)
      \end{array}
    \right)
    = e^{x H} \left(
      \begin{array}{r}
        u( 0^{\pm},t) \\
        \partial_{x } u(0^{\pm},t)
      \end{array}
    \right)
    , \pm x < 0 \label{eq:20}
  \end{equation}
  At $x=0$, the two matching conditions need be satisfied:
  \begin{eqnarray*}
    u(0^{+},t) - u(0^{-},t) =& 0 &\textrm{~(Continuity)}\\
    \partial_{x} u(0^{+},t) - \partial_{x} u(0^{-},t)  =&  -2 f(t) & \textrm{~(Differentiability)}
  \end{eqnarray*}
  For $\Im \sigma > 0$, $\lambda_{m,+}$ always has positive real part and $\lambda_{m,-}$ always has negative real part (recall \eqref{eq:defOfLambdaInAppendix}). Thus, if $u(x,t)$ is to vanish as $x \rightarrow \pm \infty$, we find that:
  \begin{eqnarray*}
    \left(
      \begin{array}{r}
        u(0^{-},t) \\
        \partial_{x } u(0^{-},t)
      \end{array} \right)
    &=& \sum_{m} u_{m,R} \phi_{m,-}(t)\\
    \left(
      \begin{array}{r}
        u(0^+,t) \\
        \partial_{x } u(0^+,t)
      \end{array} \right) &=& \sum_{m} u_{m,L} \phi_{m,+}(t)
  \end{eqnarray*}
  Since $\phi_{m,\pm}$ is a Riesz basis and $[0,f(t)] \in \hil$, we can write:
  \begin{equation}
    \label{eq:19}
    \left(
      \begin{array}{r}
        0 \\
      f(t)
    \end{array} \right)
  = \sum_{m} f_{m,+} \phi_{m,+} + f_{m,-} \phi_{m,-}
  \end{equation}
  Choosing $u_{m,R}=f_{m,-}$ and $u_{m,L}=-f_{m,+}$ solves \eqref{eq:23}, at least on a formal level. Since $u(0^{+},t) \in \hil^{+}$ and $u(0^-,t) \in \hil^{-}$, \eqref{eq:20} makes sense. Since $e^{x H}$ is bounded and analytic provided $\abs{x} < R$, this is thus an analytic mapping from $L^{2}(\storus) \rightarrow L^{2}(B_{R} \times \storus)$. 

  Now observe that both \eqref{eq:20} and \eqref{eq:19} can be analytically continued in $\sigma$, and the continuation also solves \eqref{eq:23}, therefore $G(\sigma)$ can be analytically continued in $\sigma$ as well.

  We now need only determine the behavior near $\sigma=0$. By Taylor-expanding \eqref{eq:defOfphimplusminusInAppendix} in $\sigma^{1/2}$, we find that:
  \begin{equation}
    \label{eq:12}
    \phi_{0,\pm} = N_{0} 2^{-1/2}    
    \left(
      \begin{array}{r}
        \mp \sigma^{-1/2} \\
        1
      \end{array} 
    \right) + 
    \left(
      \begin{array}{r}
        O(1) \\
        O(\sigma^{1/2})
      \end{array} 
    \right)
  \end{equation}
  while
  \begin{equation*}
    \phi_{m,\pm} = N_{m} 2^{-1/2}    
    \left(
      \begin{array}{r}
        \lambda_{m,\pm}^{-1} e^{-i m \omega t} \\
        e^{-i m \omega t}
      \end{array} 
    \right) + 
    \left(
      \begin{array}{r}
        O(1) \\
        O(\sigma^{1/2})
      \end{array} 
    \right)
  \end{equation*}
  Thus, near $\sigma=0$, we find to leading order (plugging (\ref{eq:12}) into (\ref{eq:19})) that:
  \begin{eqnarray*}
    f_{m,+} - f_{m,-} &=& 0\\
    N_{0} 2^{-1/2}(f_{m,+} + f_{m,-}) & = & P_{0} f(t)
  \end{eqnarray*}
  with $P_{0} f(t)$ projection onto the zero'th Fourier coefficient. This implies that $f_{0,\pm} = u_{0,R} = -u_{0,L} = 2^{1/2} N_{0}^{-1} (1/2)P_{0} f(t) + O(\sigma^{1/2})$. On all other coefficients, the behavior is analytic in $\sigma$ since $\lambda_{m,\pm}$ is analytic in $\sigma$ for $m \neq 0$. Thus for small $\sigma$:
  \begin{equation*}
    u(x,t) =
    (1/2)(P_{0} f(t))
    \left(
    \begin{array}{l}
      \sigma^{-1/2} \\
      1
    \end{array}
    \right)
    + O(1)
  \end{equation*}
  and therefore $G(\sigma) \delta(x) f(t) = (1/2) \sigma^{-1/2} [P_{0} f(t)] + O(1)$ near $\sigma=0$.
\end{proofof}

\section{Wellposedness}

\label{sec:wellPosedness}

Given that $Y(t)$ exists and is smooth (easily seen by using a Banach fixed point argument on \eqref{eq:floquetDuhamel}), we need to extend $Y(t)$ to $\psi_{v}(x,t)$. This is done by means of \eqref{eq:duhamelGreensFunction}; the main thing to show the extension is in $L^{2}(\mathbb{R})$ for each $t$. Since $\psi_{v,0}(x,t)$ is in $L^{2}$, we need only show that:
\begin{multline}
  \label{eq:24}
  2 i \int_0^t  \exp \left( \frac{i ( x - c ( t - s ) )^2}{4 s} \right) \psi_v ( c ( t - s ), t - s ) \frac{d s}{\sqrt{4 \pi i s}} \\
  = 2 i \int_{0}^{t} e^{i \phi(x,s,t)} Y(t-s) ds
\end{multline}
is in $L^{2}(\mathbb{R},dx)$. This is done by stationary phase. The phase of the integral (incorporating the $s^{-1/2}$ from the integrand) is $\phi(x,s,t) = (x-c(t-s))^{2}/s +i(1/2) \ln s$. The phase is stationary when:
\begin{equation*}
  \partial_{s} \phi(x,s,t) = \frac{- (x-c(t-s))^{2}}{4s^{2}}+\frac{-2(x-c(t-s))b(t-s)}{4s} - \frac{i}{2s} = 0
\end{equation*}
Provided $x > \sup_{s} \abs{c(t-s)}$, we find that $\partial_{s} \phi(x,s,t) > 0$, and thus the phase is never stationary. Further, note that $\partial_{s} \phi(x,s,t)=O(x^{2})$ near $x=\infty$. The integral over $[\epsilon,t]$ becomes:
\begin{multline}
  \label{eq:22}
  2 i \int_{\epsilon}^{t} e^{i \phi(x,s,t)} Y(t-s) ds = 2 \int_{\epsilon}^{t} i\partial_{s} \phi(x,s,t) e^{i \phi(x,s,t)} \left( \frac{Y(t-s)}{\partial_{s} \phi(x,s,t)}\right) ds\\
  = 2 \left[ e^{i \phi(x,s,t)} \left( \frac{Y(t-s)}{\partial_{s} \phi(x,s,t)}\right) \right]_{s=\epsilon}^{s=t}\\
  - 2 \int_{\epsilon}^{t} e^{i \phi(x,s,t)} \frac{
    \partial_{s} \phi(x,s,t) \partial_{s} Y(t-s) - Y(t-s) \partial_{s}^{2} \phi(x,s,t)
  }{
    \left( \partial_{s} \phi(x,s,t) \right)^{2}
  } ds
\end{multline}
Since $e^{i \phi(x,s,t)} = O(s^{-1/2})$ near $s=0$ while $\partial_{s} \phi(x,s,t)=O(s^{-2})$ near $x=0$, we find that $e^{i \phi(x,s,t)} / \partial_{s} \phi(x,s,t) \rightarrow 0$ as $s \rightarrow 0$. This implies that the first term on the right of (\ref{eq:22}) is bounded even as $\epsilon \rightarrow 0$. A simple calculation shows that $\partial_{s}^{2} \phi(x,s,t)/(\partial_{s} \phi(x,s,t))^{2}$ behaves like $s$; plugging this into the integral term of (\ref{eq:22}) shows the integrand behaves like $s^{1/2}$ near $s=0$. Since the integration is of a bounded function over a compact region, and the bounded function is $O(x^{-2})$, we find the integral decays like $x^{-2}$ as well (even as $\epsilon \rightarrow 0$).

The same argument can be applied to compute the $x$ or $t$-derivative of this; the main difference is that we must replace $Y(t-s)$ by $\partial_{x} \phi(x,s,t) Y(t-s)$ or $\partial_{t} \phi(x,s,t) Y(t-s)$, which causes the integrand in (\ref{eq:22}) to behave like $s^{-1/2}$ near $s=0$. This is still integrable\footnote{We can not repeat this trick any further. If repeated once more, we  obtain terms behaving like $s^{-3/2}$ near $s=0$, which is not integrable.}. 

{\bf Acknowledgements: }
  We thank A. Soffer and M. Kiessling for useful discussions. J.L.L. and O.C. would like to thank the IHES in Bures-sur-Yvette and the IAS in Princeton where part of the work was done. We also thank an anonymous referee for a very careful reading. Work supported by NSF Grants DMS-0100495, DMS-0406193, DMS-0600369, 
  DMS01-00490, 
  DMR 01-279-26 and AFOSR grant AF-FA9550-04. 
  Any opinions, findings, conclusions or recommendations expressed in this material are those of the authors and do not necessarily reflect the views of the National Science Foundation.

\nocite{MR1664792}
\nocite{MR1868995}
\bibliographystyle{plain} \bibliography{stucchio}

\begin{figure}
  \begin{center}
    \includegraphics[scale=0.7]{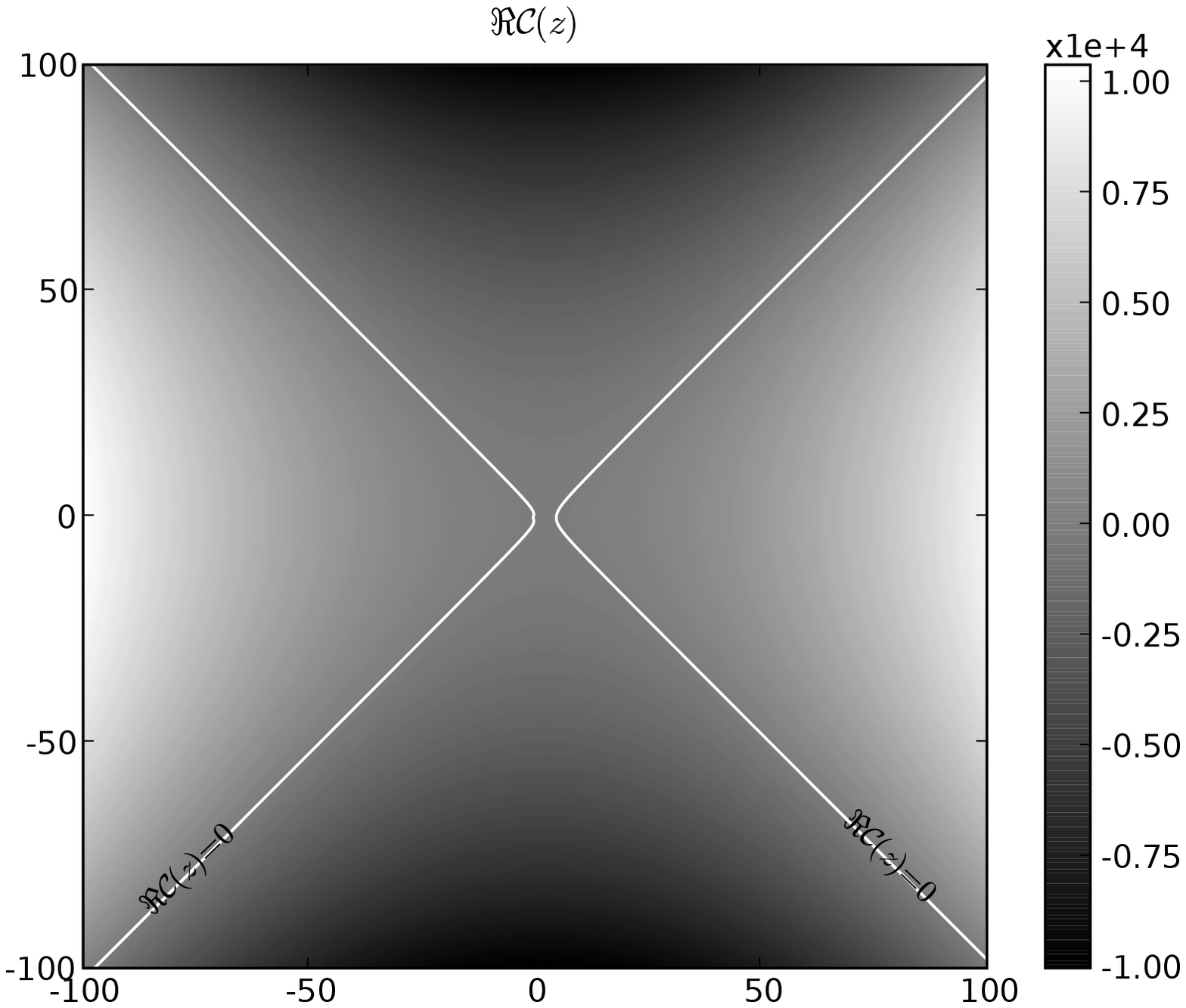}
  \end{center}
  \caption{A plot of $\Re \mathfrak{C}(z)$, for the specific choice of $\mathfrak{C}(z)=z^{2}-5z-5z^{-1}+z^{-2}$. As is apparent from the figure, the regions $\Re \mathfrak{C}(z) > 0$ and $\Re \mathfrak{C}(z) < 0$ are approximately sectors for sufficiently large $\abs{z}$.}
  \label{sector diagram}
\end{figure}

\end{document}